\documentclass{iopart}
\usepackage{iopams}
\usepackage{braket,amsmath,color,graphicx,cite}
\usepackage[margin=2cm]{geometry}
\usepackage[colorlinks,linkcolor={blue},citecolor={blue}]{hyperref}
\usepackage{mathrsfs}

\newcommand{\pin}{\par\noindent}
\allowdisplaybreaks

\begin{document}

\title{Universal entanglement signatures of quantum liquids as a guide to fermionic criticality}

\author{Siddhartha Patra$^{1,*}$, Anirban Mukherjee$^{1}$ and Siddhartha Lal$^{1}$}
\eads{\mailto{siddharthaptr@gmail.com}, \mailto{amukh1290@gmail.com}, \mailto{slal@iiserkol.ac.in}}

\address{$^1$Department of Physical Sciences, Indian Institute of Science Education and Research-Kolkata, W.B. 741246, India}
\address{$^*$Author to whom any correspondence should be addressed.}

\date{\today}

\begin{abstract}

\pin 
An outstanding challenge involves understanding the many-particle entanglement of liquid states of quantum matter that arise in systems of interacting electrons. The Fermi liquid (FL) shows a violation of the area-law in real-space entanglement entropy of a subsystem, believed to be a signature of the ground state of a gapless quantum critical system of interacting fermions. Here, we apply a $T=0$ renormalisation group approach to the FL, unveiling the long-wavelength quantum fluctuations from which long-range entanglement arises. A similar analysis of non-Fermi liquids such as the 2D marginal Fermi liquid (MFL) and the 1D Tomonaga-Luttinger liquid (TLL) reveals a universal logarithmic violation of the area-law in gapless electronic liquids, with a proportionality constant that depends on the nature of the underlying Fermi surface. We extend this analysis to classify the gapped quantum liquids emergent from the destabilisation of the Fermi surface by RG relevant quantum fluctuations arising from backscattering processes.

\end{abstract}
\maketitle

\section{Introduction}
\par\noindent 
There has been a surge recently in applying quantum information-theoretic tools towards understanding the nature of entanglement and quantum correlation encoded within the ground states and low-lying excitations of strongly correlated systems~\cite{entanglement_1,eisert2010colloquium,Laflorencie2016}. For a system with local interactions among its constituents, the entanglement of a subsystem is expected to follow an area-law (i.e., scale with the size of its boundary)~\cite{eisert2010colloquium}. There are, however, some notable departures. First, a gapped topologically ordered system possesses, in addition to the area-law term, a sub-dominant piece that arises from its topological properties and encodes true long-range entanglement~\cite{Levin_Wen_2005,Kitaev_Preskill_2006,Hamma_Zanardi_2005}. Another exception involves the
finding of signatures of volume-law entanglement in some quantum critical systems arising from the existence of gapless quantum fluctuations at all lengthscales~\cite{volumelaw_1,volumelaw_2,volumelaw_3,wang2013constructing,peng2021gapless}. A final exception is found in gapless quantum liquids of interacting electrons in spatial dimensions $d>1$, and forms the focus of our work. The best understood gapless electronic liquids are metals belonging to the Fermi liquid paradigm~\cite{JETP5,JETP6,JETP7}. 
These quantum critical systems are known to possess long-range entanglement in the form a modified area-law: $S_{EE} \sim L^{d-1} \ln L$ for $d$-dimensional subsystem of spatial length $L$~\cite{EE_fermi_2,EE_fermi_4,EE_metal_vio_1,EE_metal_vio_2,
EE_metal_vio_3,EE_metal_vio_4,EE_metal_vio_5}. The extra $\ln L$ is conjectured to arise from the presence of gapless long-wavelength quasiparticle excitations lying 
proximate to a Fermi surface~\cite{EE_fermi_1,EE_fermi_2,EE_fermi_3,EE_fermi_4,EE_fermi_5}. Indeed, this result is believed to be the higher dimensional analogue of the entanglement scaling observed in quantum critical phases of interacting quantum spin and electron systems in 1D:~$S_{EE}^{1D}\sim \ln L$~\cite{EE_Cr_1,EE_Cr_2,EE_Cr_3,EE_Cr_4}. A similar violation of the area-law is also expected in other quantities of such gapless liquids, such as the fluctuation in the number of particles within a subsystem~\cite{EE_fluc_1,EE_fluc_2,EE_fluc_3,EE_fluc_4,EE_fluc_5,EE_fluc_6,EE_fluc_7,EE_fluc_8,EE_fluc_9,
EE_fluc_10,EE_fluc_11,EE_fluc_12,EE_fluc_13,EE_fluc_14,EE_fluc_15,EE_fluc_16}. 
\pin
In addition to the FL, we will also focus here on a particular variant of the 2D non-Fermi liquid known as the marginal Fermi liquid (MFL). 
Proposed on phenomenological grounds as the parent metal of high-temperature superconductivity in the cuprates~\cite{varma-physrevlett.63.1996,MFL_2,MFL_4}, a first-principles derivation of the effective theory for the MFL (e.g., low-energy Hamiltonian, nature of gapless excitations etc.) was obtained only recently from a detailed renormalisation group study of the 2D Hubbard model~
\cite{MukherjeeMott1,MukherjeeMott2,MukherjeeNPB2}. 
Interestingly, the modified area-law discussed above for the real-space entanglement entropy of the FL is also proposed to hold for non-Fermi liquids such as the MFL~\cite{EE_fermi_num_10}, hinting at a possible universality connecting various gapless electronic quantum liquids despite the obvious differences in some of their key properties (e.g., nature of their low-energy excitations etc.).

\pin
\textbf{\textit{Gapless electronic liquids are different.}}~A deeper understanding of the entanglement properties of gapless quantum liquids such as the FL and MFL necessitates an approach that lies beyond the well-established Ginzburg-Landau-Wilson (GLW) paradigm. This is can be argued as follows: the effective theories for phases falling within the GLW paradigm describe the quantised excitations of classical scalar (or bosonic) field degrees of freedom that lie above symmetry broken ground states that are short-range entangled at best. They are characterised by order parameters that correspond to the ground state expectation values of real-space local operators, and correlation functions that satisfy the cluster decomposition property. On the other hand, the excitations of a system of interacting fermions carry sign factors that arise from the exchange of particles and do not have a classical origin. Further, a filled Fermi volume at $T=0$ is the ground state for gapless fermionic quantum liquids, owes its origin to the Pauli exclusion principle, and is characterised by a topological quantum number called the Luttinger volume that is robust against the inclusion certain kinds of inter-particle interactions~\cite{luttinger1960ground,oshikawa2000topological,dzyaloshinskii2003some,seki2017topological, Heath2020,MukherjeeNPB1,MukherjeeTLL}. Indeed, these liquids are described by 
effective theories that contain only the physics of the low-energy long-wavelength degrees of freedom 
proximate to the Fermi surface~\cite{shankar1994,polchinski1992}. The remarkable simplicity of these theories is that they are comprised 
purely of terms that are number-diagonal in the momentum (or, $k$)-space single-electron Fock state occupation 
number operators $\hat{n}_{\vec{k}}$ (see equations eq.\eqref{eq:rgfixed_fl} and eq.\eqref{eq:rgfixed_mfl} below). This indicates 
that the $k$-space many-particle wavefunctions of these quantum liquids are direct product in form, 
i.e., they are separable in terms of the single-electron Fock states and do not encode any entanglement 
among them. Instead, as mentioned above, these states of electronic quantum matter display long-range 
entanglement in real-space. From a renormalisation group (RG) perspective, these low-energy (IR) 
theories correspond to universal scale-invariant quantum critical fixed points obtained from the 
coarse-graining of bare (UV) theories of interacting electrons possessing translation invariance. 
\par\noindent
\textbf{\textit{Setting out the goals.}}~ 
We first seek the quantum fluctuations in the UV from which the real-space long-range entanglement of these universal IR theories is emergent. As the low-energy degrees of 
freedom in the IR are 
effectively decoupled from their UV counterparts, meeting this challenge necessitates 
an investigation of the forward and tangential scattering related RG relevant quantum fluctuations that are resolved under the flow from UV to 
IR. In keeping with the Wilsonian approach to critical phenomena, this involves understanding the 
signatures of universality that are likely encoded within the many-particle entanglement wrought from 
such quantum fluctuations. The answer to this question holds the potential to offer crucial insight on 
whether certain aspects of the physics of candidate non-Fermi liquids (e.g., the marginal Fermi liquid) 
show commonalities with metals belonging to the Fermi liquid paradigm, even if some other aspects (e.g., 
the nature of the low-lying excitations) are qualitatively different. To the best our knowledge, such a study of the RG evolution of the $k$-space entanglement has been attempted only for the case of scalar (or bosonic) field theories~\cite{balasubramanian2012momentum}.
\pin
Systems of interacting fermions, 
on the other hand, involve constraints on the inter-particle scattering mechanisms arising from 
the Pauli exclusion principle as well as phase space constraints due to the existence of a well-defined 
Fermi volume and a bounding Fermi surface~\cite{shankar1994,polchinski1992}. We recall that the quantum fluctuations in such systems thus contain fermion-exchange related signs, rendering their study difficult. Recent studies indicate that the sign structure of the many-particle wavefunction significantly affect the nature of the entanglement encoded in them~\cite{grover2014,grover2015}. This is of particular significance to the study of non-Fermi liquid gapless states that are emergent at quantum critical points related to the collapse of a Fermi liquid metal: a recent numerical study based 
on an ansatz wavefunction incorporating correlations from long-ranged backflow effects of the interacting fermions revealed a crossover from a volume-law scaling of the subsystem entanglement entropy to the modified area-law familiar for the Fermi liquid effective theory~\cite{kaplis2017}. The crossover takes place across a distance scale related to the physics of critical backflow interactions. In addition, it is pertinent to 
enquire on whether there exist similar imprints of universality hidden within the backscattering related RG relevant quantum 
fluctuations that destabilise the Fermi surface of these gapless metallic states and lead to the 
emergence of topologically ordered gapped liquid states of quantum matter (e.g., the 2D Mott liquid found in the 2D Hubbard model on the square lattice at $1/2$-filling~\cite{MukherjeeMott1} and the Cooper pair insulator for the reduced BCS model with a circular Fermi surface~\cite{siddharthacpi}). This amounts to an attempt at a systematic classification of the 
states of quantum matter based on their entanglement content, and looks beyond the short range entangled 
states that belong to the Ginzburg-Landau-Wilson paradigm of broken symmetries and real-space local order 
parameters.  
\par\noindent
\textbf{\textit{The strategy.}}~In order to meet these goals, we adopt a RG approach that we have formulated recently for the 
investigation of criticality observed in systems of interacting fermions~\cite{MukherjeeMott1,MukherjeeMott2,pal2019,MukherjeeNPB1,
MukherjeeNPB2,MukherjeeTLL,siddharthacpi,kondo_urg,patramck2022}. Formulated using only many-particle unitary transformations to the bare Hamiltonian, the procedure reveals IR fixed point theories and their low-energy wavefunctions. An 
iterative application of the unitaries to the IR wavefunctions then generates a family of states spanning 
towards the UV. This facilitates a quantitative study of the RG evolution of many-body correlations, as well as several many-particle entanglement features, from the ground state and lowest lying excited states~\cite{MukherjeeTLL,siddharthacpi,kondo_urg,MukherjeeHolography,patramck2022}. 
Thus, we first employ this strategy to obtain the UV wavefunctions of several gapless and gapped quantum liquid states that are unitarily connected to their IR counterparts. We then 
compute the scaling of the entanglement entropy of a block of states in $k$-space (lying proximate to the IR Fermi energy) with the block size ($\Lambda$). 
\pin
We note that there is a small, but growing, body of work on adopting a field theoretic approach towards understanding the dependence of many-particle entanglement on scale in systems with inter-particle interactions. While most of these works apply the Wilsonian renormalisation group towards obtaining entanglement measures from the effective action of interacting scalar theories in real-space~\cite{miqueleto2021,klco2021,iso2021I,iso2021II,nishiokaRMP} and in momentum-space~\cite{balasubramanian2012momentum,cesar2018,agon2018,kawamoto2021,grignani2017,
peschanski2016,lundgren2014,lundgren2019,balasubramanian2013,guijosa2022,costa2022,costa2022arXiv}, some recent works along these lines have also been devoted to the study of interacting fermions at finite densities in momentum-space~\cite{mcdermott2013,flynn2022}. Our work, on the other hand, takes the Hamiltonian renormalisation route in momentum-space towards the same goal. As discussed in the concluding section, our methods corresponds to a tensor network that readily admits an exact holographic mapping.   

\section{Methods}
We now present the main methods employed by us. As these methods have been employed by us in several recent works~\cite{MukherjeeMott1,MukherjeeMott2,pal2019,MukherjeeNPB1,
MukherjeeNPB2,MukherjeeTLL,siddharthacpi,kondo_urg,MukherjeeHolography,patramck2022}, we present only an overview here and refer the reader to \ref{URGdetails} as well as our earlier works for further details.
\subsection{Unitary renormalization group (URG) method}
\pin
The URG method iteratively decouples quantum fluctuations in higher energy Fock states by applying a many-body unitary operator, generating thereby a low-energy effective Hamiltonian. For a general electronic system with Hamiltonian $\mathcal{H}_{(N)}$ with total $N$ Fock states, Mutually non-commuting terms in the Hamiltonian are the source of quantum fluctuation in the occupation of the electronic Fock states. We first label the Fock states according to the eigenvalues of the diagonal part of the bare Hamiltonian, $\epsilon_{N}>\epsilon_{N-1}>\cdots$. The unitary operator $\mathcal{U}_{N}$ then decoupled the $N^{th}$ Fock state from all other states, i.e., it removes quantum fluctuation present in this state. The low-energy effective Hamiltonian after the first step of URG is 
\begin{eqnarray}
\mathcal{H}_{(N-1)}=\mathcal{U}_{N} \mathcal{H}_{(N)} \mathcal{U}_{N}^{\dagger}~.
\end{eqnarray}
Subsequently, the unitary operator $\mathcal{U}_{N-1}$ is used to remove the quantum fluctuations in the Fock state $N-1$, and so on. The general form of the unitary operator for $j^{th}$ step of URG is given as
\begin{eqnarray}
\mathcal{U}_{j} &=&(1+\eta_{j}-\eta_{j}^{\dagger} ) /\sqrt{2}\quad,\quad 
\eta_{j}^{\dagger} = \frac{1}{\omega-\textrm{Tr}(\mathcal{H}_{(j)} \hat{n}_j )} c_j^{\dagger} \textrm{Tr} (\mathcal{H}_{(j)} c_j)~,
\end{eqnarray}
where $\omega$ characterises the energyscale for quantum fluctuations,
and an equivalence of $\omega$ has been established with that for thermal fluctuations at a finite-temperature~\cite{MukherjeeMott1,MukherjeeMott2,MukherjeeNPB1,MukherjeeNPB2}. Iterative application of the unitaries generates a series of effective Hamiltonians with progressively lower RG energy scale, $\mathcal{H}_{(N)},\mathcal{H}_{(N-1)},\cdots$. The fixed point effective theory is reached at the effective Hamiltonian $\mathcal{H}_{(j^*)}$, when further decoupling is not possible due to the vanishing of the denominator of $\eta_{j^*}$.

\par\noindent 
At any step of the URG~\cite{pal2019,MukherjeeMott1,MukherjeeMott2,MukherjeeNPB1,MukherjeeNPB2,MukherjeeTLL,
MukherjeeHolography,siddharthacpi}, the Hamiltonian in the rotated basis is given by
\begin{eqnarray}
U_{k\sigma}\hat{H} U_{k\sigma}^{\dagger} &&= \frac{1}{2}Tr_{k\sigma}(\hat{H})+\tau_{k\sigma}Tr_{k\sigma}(H \tau_{k\sigma})
+ \tau_{k\sigma}\{c_{k\sigma}^{\dagger} Tr_{k\sigma}(\hat{H} c_{k\sigma}),\hat{\eta}_{k\sigma} \}~.
\end{eqnarray}
It is important to note that while $\hat{n}_{k\sigma} \hat{H} (1-\hat{n}_{k\sigma}) \neq 0$ (i.e., there exist non-trivial quantum fluctuations in the occupation of single-particle Fock state given by $n_{\sigma}$) prior to the application of the unitary operator, the unitary operation removes such quantum fluctuation.
\begin{eqnarray}
\hat{n}_{k\sigma} U_{k\sigma} \hat{H} U_{k\sigma}^{\dagger} (1-\hat{n}_{k\sigma})=0~\Rightarrow~[\hat{n}_{k\sigma},U_{k\sigma} \hat{H} U_{k\sigma}^{\dagger}]=0~,~~
\label{rg_condition}
\end{eqnarray}
upon the application of $U$. The degree of freedom $n_{k\sigma}$ is thus rendered an integral of motion (IOM) of the RG flow. The RG equations can then be obtained from the condition eq.\eqref{rg_condition}. The diagonal part of the Hamiltonian $H^D_{(k\sigma)}$ with respect to the Fock state ($k\sigma$) has the property $[H^D_{(k\sigma)},\hat{n}_{k\sigma}]=0$. Then, the operators $\eta_{k\sigma}^{\dagger}$ and $\eta_{k\sigma}$ are given as
\begin{eqnarray}
\eta_{k\sigma}^{\dagger} &=& \frac{1}{\hat{\omega}_{k\sigma}-Tr_N(\hat{n}_{k\sigma} H^D_{(k\sigma)})\hat{n}_{k\sigma}}  \mathcal{T}_{k\sigma}^{\dagger}\quad,\quad 
\eta_{k\sigma} = \frac{1}{\hat{\omega}_{k\sigma}-Tr_{k\sigma}((1-\hat{n}_{k\sigma}) H^D_{(k\sigma)})(1-\hat{n}_{k\sigma})} \mathcal{T}_{k\sigma} ~,
\end{eqnarray}
\pin 
where $\mathcal{T}_{k\sigma}^{\dagger}=Tr_{k\sigma} (H_{(k\sigma)} c_{k\sigma})c_{k\sigma}^{\dagger}$, and $\mathcal{T}_{k\sigma}=c_{k\sigma} Tr_{k\sigma} (c_{k\sigma}^{\dagger}H_{(k\sigma)} )$. Such decoupling leads to the emergence of a lower energy effective Hamiltonian $U_{k\sigma} \hat{H} U^{\dagger}_{k\sigma}$. Further iterative application of unitary operators on effective Hamiltonians leads to the decoupling of further higher energy (UV) Fock states and the emergence of lower energy effective Hamiltonians. Importantly, an equivalence of the quantum fluctuation scale $\omega$ with an energyscale for finite-temperature thermal fluctuations has also been established in Refs.\cite{MukherjeeMott1,MukherjeeMott2,MukherjeeNPB1,MukherjeeNPB2}. RG flow equations for various couplings are derived from the series of effective Hamiltonians. The RG flow attains a stable effective Hamiltonian at the IR fixed point 
when further decoupling is not possible. 
The URG has been applied in deriving analytic expressions for the low-energy effective theories of several gapless (e.g., Fermi liquid, Marginal Fermi liquid, Tomonaga-Luttinger liquid) as well as gapped quantum liquid states of matter discussed in the introduction (e.g., Mott liquid, Cooper Pair Insulator, etc.)~\cite{MukherjeeMott1,MukherjeeMott2,pal2019,MukherjeeNPB1,
MukherjeeNPB2,MukherjeeTLL,siddharthacpi}. 
We will next see that the URG motivates a renormalisation group scheme for the study of many-particle entanglement in these quantum liquids.


\subsection{Momentum-space entanglement renormalisation group (MERG)}

\pin 
Having set out the URG flow to the IR fixed point Hamiltonian, we can now sketch the MERG algorithm by which to study the renormalisation group flow of entanglement in momentum-space. First, we solve for the wavefunctions of the ground state and low-lying excited states of the fixed point Hamiltonian. Now, by acting with the unitary operators of the URG in reverse on, say, the ground state $|\Psi_0\rangle$, we systematically re-entangle the low-energy degrees of freedom lying proximate to the Fermi surface with the hitherto decoupled Fock states (i.e., describing the IOMs of the URG flow). In this way, we generate under the MERG flow a family of many-body wavefunctions that are unitarily connected to $|\Psi_0\rangle$ by quantum fluctuations in momentum-space~\cite{MukherjeeTLL,siddharthacpi,kondo_urg,MukherjeeHolography}. 
The explicit form of the wavefunction obtained after the first step of the MERG flow is given by 
\begin{eqnarray}
|\Psi_{1}\rangle &=&  \frac{1}{\sqrt{2}} \bigg[ 1+ Q_{k\sigma} c_{k\sigma} Tr_{k\sigma} (c_{k\sigma}^{\dagger}H_{(k\sigma)} )
- P_{k\sigma} Tr_{k\sigma} (H_{(k\sigma)} c_{k\sigma})c_{k\sigma}^{\dagger}  \bigg] |\Psi_0\rangle~, \nonumber\\
\hat{Q}_{k\sigma}&=&  \frac{1}{\hat{\omega}_{k\sigma}-Tr_{k\sigma}((1-\hat{n}_{k\sigma}) H^D_{(k\sigma)})(1-\hat{n}_{k\sigma})}\quad,\quad  
\hat{P}_{k\sigma}= \frac{1}{\hat{\omega}_{k\sigma}-Tr_{k\sigma}(\hat{n}_{k\sigma} H^D_{(k\sigma)})\hat{n}_{k\sigma}} ~.
\label{eq:reverseRG-state}
\end{eqnarray}
Having obtained a series of many-body wavefunctions from the MERG, we compute from them the RG evolution of several measures of many-particle entanglement and many-body correlations.

\section{Models of quantum matter}
We introduce the various states of quantum matter whose entanglement has been studied using the methods outlined above. In order to focus the discussion, we present all technical details (e.g., the derivation of URG flow equations that lead to various IR fixed point theories) in \ref{URGdetails}.  
\subsection{Gapless quantum liquids}
We investigate the momentum-space entanglement features of a variety of gapless states of fermionic quantum matter in this work, each of which is described briefly below.
\par\noindent\textbf{1. Fermi liquid and Marginal Fermi liquid for the square lattice:}\\
The IR fixed point Hamiltonian describing the excitations of the FL phase is~\cite{JETP5,JETP6,JETP7}
\begin{eqnarray}
H^{\hat{s}}_{FL}&=& \displaystyle\sum_{k,\sigma} \epsilon_{k} n_{k\sigma} +\displaystyle\sum_{ \substack{kk'\in \{\hat{s},\hat{s}'\}\\ \sigma,\sigma'}} V_{\substack{kk'\sigma\sigma'}}~  n_{k,\sigma} n_{k',\sigma'}~,
\label{eq:rgfixed_fl}
\end{eqnarray}
where $\hat{s}$ and $\hat{s}'$ correspond to excitations in directions normal to the Fermi surface, and 
an analytic expression for the RG evolution of the interaction coupling $V_{\substack{kk'\sigma\sigma'}}$ has been obtained from a unitary renormalisation group (URG) study of the 2D Hubbard model on the square lattice at large hole doping~\cite{MukherjeeMott2,MukherjeeNPB2}. 
Further, the IR fixed point effective Hamiltonian for the MFL phase is obtained from the URG analysis of the 2D Hubbard model at half-filling, as well as optimal doping, as as~\cite{MukherjeeMott1,MukherjeeMott2,MukherjeeNPB2}
\begin{eqnarray}
H^{\hat{s}}_{MFL}&=&\displaystyle\sum_{k,\sigma} \epsilon_{k} n_{k\sigma} + \hspace*{-0.5cm}\displaystyle\sum_{ \substack{\tilde{K}=(kk'k''\sigma) \\kk'k''\in \hat{s}\\\sigma}} \hspace*{-0.5cm}\mathcal{R}^{\hat{s}}_{\tilde{K}} n_{k,\sigma} n_{k',-\sigma} \bigg(1-n_{k''\sigma}\bigg)~,
\label{eq:rgfixed_mfl}
\end{eqnarray}
where the RG evolution for the interaction coupling $\mathcal{R}^{\hat{s}}_{\tilde{K}}$ shows the emergence of an excitation different from the Fermi liquid kind. We can see from eq.\eqref{eq:rgfixed_mfl} that the low-lying excitations of the MFL are described by three-particle composite objects comprised of two electrons and a hole. Further, various features conjectured for the MFL phenomenology (e.g., the vanishing of the Landau quasiparticle residue at energies proximate to the Fermi surface~\cite{varma-physrevlett.63.1996,MFL_2,MFL_4}) have been established from the MFL effective Hamiltonian eq.\eqref{eq:rgfixed_mfl}~\cite{MukherjeeMott1,MukherjeeMott2,MukherjeeNPB2}. Importantly, all backscattering related quantum fluctuations in $k$-space are rendered irrelevant under the RG flows to the FL and MFL fixed point theories. On the other hand, quantum fluctuations arising from RG relevant forward and tangential scattering processes of interacting electrons are of interest in understanding the nature of the entanglement. Finally, in unveiling the entanglement features of the gapped quantum liquids (e.g., Mott liquid, Cooper pair insulator) obtained by destabilising the FL and MFL fixed points, we will consider the quantum fluctuations arising from RG relevant backscattering processes. 
\begin{figure}
\centering
\includegraphics[scale=1]{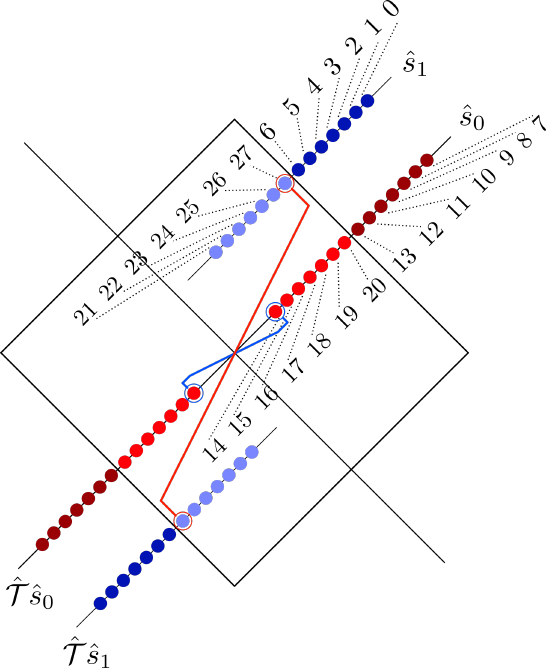}
\caption{
\textbf{Momentum-space construction of reduced system studied under MERG.} Four directions ($\hat{s}_0,\hat{s}_1,\mathcal{\hat{T}}\hat{s}_0,\mathcal{\hat{T}}\hat{s}_1$) are chosen normal to the Fermi surface in the 2D Brillouin zone; two ($\hat{s}_0,\mathcal{\hat{T}}\hat{s}_0$, lines of red circles) are labelled as the nodal directions, and two ($\hat{s}_1,\mathcal{\hat{T}}\hat{s}_1$, lines of blue circles) as the near-anti-nodal. Each $\hat{s}$-direction is comprised of 28 electronic Fock states (red and blue circles). Pairs of electronic Fock states with opposite momentum and spin ($\vec{k}\uparrow,-\vec{k}\downarrow$) form Anderson pseudospins. The bold blue and red lines show the two opposite momentum present at two ends, participating in the pseudospin formation.
}
\label{fig:cooperchannel}
\end{figure}


\par\noindent\textbf{2. Fermi liquid with circular Fermi surface in 2D:}\\ 
In contrast to the electronic differentiation in the electronic dispersion that is present for the case of the FL and MFL phases on the square lattice, the FL phase for the case of a circular Fermi surface is isotropic in the angular direction.
Thus, in order to understand the dependence of various entanglement and scattering signatures of the FL phase on the electronic differentiation, we have also studied the momentum space entanglement scaling of an FL phase proximate to a circular Fermi surface~\cite{siddharthacpi}. 
Though the nature of the forward scattering for both cases is similar, the difference in the nature of tangential scatterings makes them different. Further, the symmetry of the circular Fermi surface allows one to focus on the momentum space scaling for any one radial direction; this renders possible the simulation of a larger number of steps of the MERG, and thereby obtain a larger family of wavefunctions ranging between the IR and UV.
\par\noindent\textbf{3. Tomonaga-Luttinger Liquid in 1D:}\\ 
In order to understand the dependence of the nature of the many-particle entanglement on the dimensionality of the underlying electronic system, we have studied the gapless quantum liquid phase in one spatial dimension known as the Tomonaga-Luttinger Liquid (TLL)~\cite{MukherjeeTLL}. While this phase  corresponds to a non-Fermi liquid, the logarithmic entanglement scaling of this phase in real-space~\cite{EE_Cr_4,EE_fluc_15} hints at an underlying universality across dimensions. Due to the simplicity of the physics being confined to one spatial dimension, we can once again simulate a larger family of wavefunctions from the MERG flow.
\par\noindent\textbf{4. Quantum fluctuations proximate to a quantum critical point:}\\ 
An important class of gapless quantum matter are those found at quantum critical points. Here, we study the entanglement features of the gapless quantum liquid obtained at the quantum critical point (QCP) obtained recently by some of us at optimal hole-doping in the 2D Hubbard model~\cite{MukherjeeNPB1, MukherjeeNPB2}. The four nodal directions of the Fermi surface were observed to be gapless at this critical point, whereas other directions perpendicular to the Fermi surface were found to be gapped (and comprised of the 2D Mott liquid). We expect a volume law behaviour for the entanglement entropy arising from the strong quantum fluctuations present at such a QCP. 
\subsection{Gapped quantum liquids}
In order to contrast the entanglement features of various gapless quantum liquids with their gapped counterparts, we have performed MERG calculations of two examples of the latter as well. 
\par\noindent\textbf{1. 2D Cooper Pair Insulator:}\\  
The Cooper pair insulator is an insulating phase emergent from a RG relevant pairing instability of the FL phase. The momentum space symmetry of the emergent CPI phase ($C_4$ or $U(1)$)~\cite{MukherjeeMott2,MukherjeeNPB2,siddharthacpi} corresponds to the symmetry of the Fermi surface of the parent metallic phase (optimally hole-doped Fermi surface of the square lattice and circular Fermi surface respectively).
\par\noindent\textbf{2. 2D Mott liquid:}\\ 
The 2D Mott liquid phase is an insulating phase emergent from an instability of the Marginal Fermi liquid phase of the 2D Hubbard model on the square lattice at $1/2$-filling. It arises from RG relevant backscattering of charge pseudospin degrees of freedom~\cite{MukherjeeMott1,MukherjeeNPB2,MukherjeeHolography}, and possesses the $C_{4}$ symmetry of the square Fermi surface of a half-filled tight-binding problem on the square lattice. 


\section{Results}

\pin
\textbf{\textit{Forming a reduced system}.}
In order to study the $T=0$ FL and MFL phases obtained recently for the Hubbard model ~\cite{MukherjeeNPB1,MukherjeeNPB2}, we consider the square Fermi volume of the $1/2$-filled tight-binding problem on the square lattice and divide the momentum-space window of low-momenta around the square Fermi surface into several directions normal to it (see Fig.\ref{fig:cooperchannel}). We note that electrons among all the directions perpendicular to the Fermi surface interact with one other, and will therefore contribute to the renormalization group computation. We consider scattering processes among all such electronic states in implementing the URG and MERG formalisms. 

Our investigations of entanglement involve simulating the RG evolution of many-body wavefunctions. For this, we apply iteratively the inverse many-particle unitary transformations of the URG method to the IR ground state wavefunction of, say, the FL (see Method). This generates quantum fluctuations by coupling the electronic Fock states near the Fermi surface to those farther away via forward and tangential scattering processes. Note that any two-particle scattering process can always be decomposed in forward, backward, and tangential scattering. In this way, we obtain a family of wavefunctions ranging from IR to UV, and involving a systematic growth of the size of the effective Hilbert space of interacting electrons. This $k$-space entanglement renormalisation group (MERG) process is, however, rendered very challenging due to the fact that the many-particle Hilbert space scales exponentially in the system size. In order to make the MERG analysis tractable, we construct a reduced subspace of the FL and MFL systems in $k$-space for an underlying 2D square lattice (Fig.\ref{fig:cooperchannel}). We consider only the quantum fluctuations generated by scattering processes along only four such directions ($\hat{s}_0,\hat{s}_1,\mathcal{\hat{T}} \hat{s}_0, \mathcal{\hat{T}} \hat{s}_1$). The direction $\hat{s}_{0}$ and its inverse ($\mathcal{\hat{T}}\hat{s}_{0}$) are along the nodal direction of the square Brillouin zone, while $\hat{s}_1$ and its inversion ($\mathcal{\hat{T}} \hat{s}_1$) are near the anti-nodal direction.
The Fock states residing on the four $\hat{s}-$directions considered form a minimal set that captures all the possible (forward, backward, and tangential) scattering processes. We have used this construction to capture difference and universality among various gapless and gapped quantum liquids.

\begin{figure}[!htb]
\centering
\includegraphics[width=0.35\textwidth]{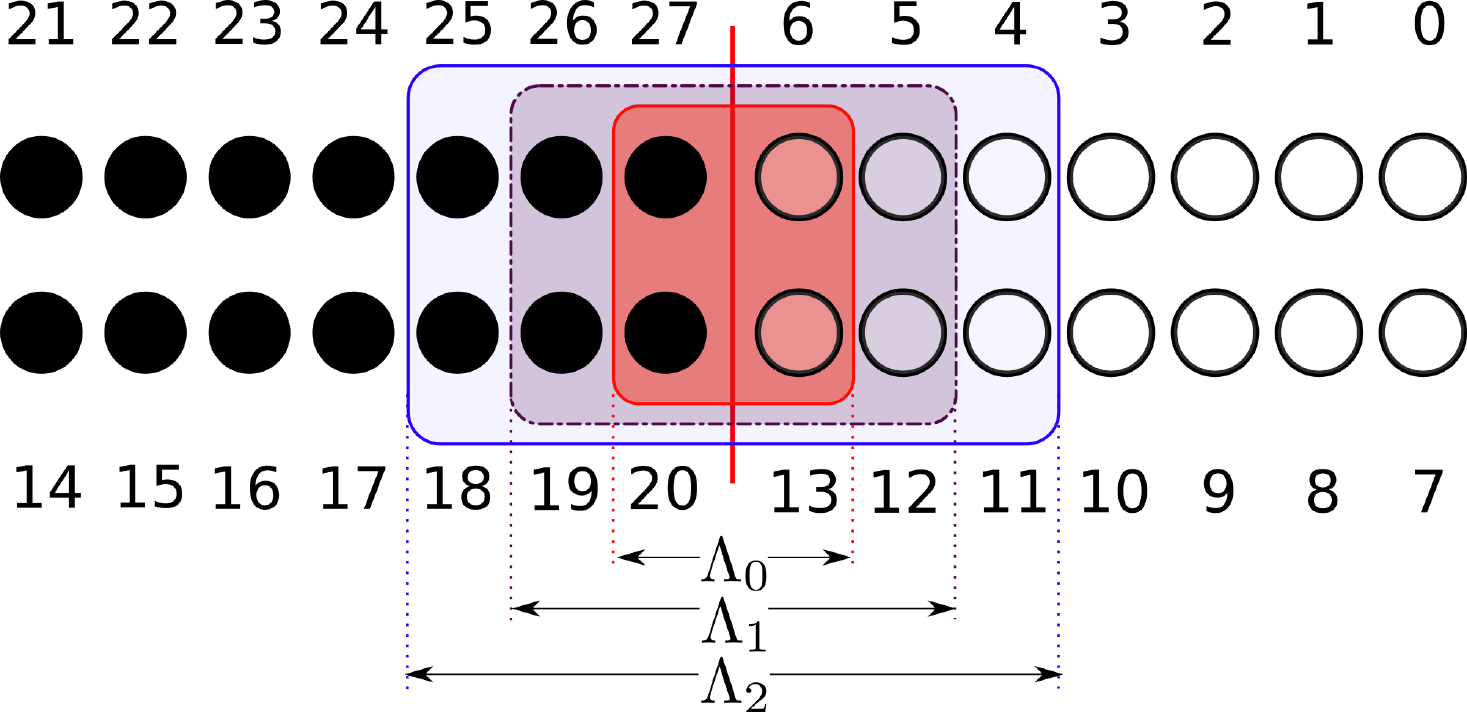}
\caption{\textbf{Progression of the MERG flow.} The black circles represent up-pseudospins (pairs of occupied electronic Fock states), and the white circles represent down-pseudospins (pairs of unoccupied electronic Fock states). The red vertical line represents the Fermi surface. The red shaded box represents the $k$-space window ($\Lambda_{0}$) of the IR fixed point theory, while the light purple and light blue shaded boxes represent the windows ($\Lambda_{1}$ and $\Lambda_{2}$) of electronic Fock states coupled by quantum fluctuations after the first and second steps of the MERG respectively. 
}
\label{fig:RGwindow}
\end{figure}
\pin 
As a study of the 112 fermionic states (i.e., a Hilbert space dimension of $\sim 10^{33}$) is intractable, 
we focus the study on the quantum fluctuations of the holon (pairs of unoccupied electronic Fock states) and doublon (pairs of occupied electronic Fock states) degrees of freedom in the hole-doped 2D Hubbard model. As shown in Fig.\ref{fig:cooperchannel}, this corresponds to a smaller sub-space defined in terms $56$ pseudospin degrees of freedom (defined in terms of electronic Fock states with opposite momenta) and a reduced Hilbert space size $\sim 10^{16}$~\cite{anderson1958random,MukherjeeMott1,MukherjeeMott2,MukherjeeHolography}. 
A further drastic simplification is made by ignoring the RG irrelevant backscattering related quantum fluctuations: this allows us to study
a reduced system of 28 pseudospins residing on only one side of the Brillouin zone, say, the $\hat{s}_{0}$ and $\hat{s}_{1}$ directions. 
In each of these directions, we have chosen 14 pseudospin states, seven outside and seven inside the Fermi surface. We have labelled all 28 pseudo-spins using integers in Fig.\ref{fig:RGwindow}. Specifically, Fock states residing outside or inside the Fermi surface along a particular $\hat{s}-$direction are labelled such that states nearest/farthest to the Fermi surface are associated with highest/lowest integer respectively. Further, along a particular $\hat{s}-$direction, states inside the Fermi surface are labelled by a higher integer value than those outside the Fermi surface. 

The MERG proceeds via the iterative application of a set of unitaries ($U_{(j)}^{\dagger}~,~j\in\{1,2,\cdots,6\}$) to the IR fixed point ground state and lowest-lying excited state wavefunctions of the gapless FL and MFL liquids that systematically couple the $4$ electronic Fock states ($6,13,20,27$) lying within the window proximate to the Fermi surface to the $24$ Fock states lying outside. 


\begin{figure}[!htb]\centering
\includegraphics[width=0.48\textwidth]{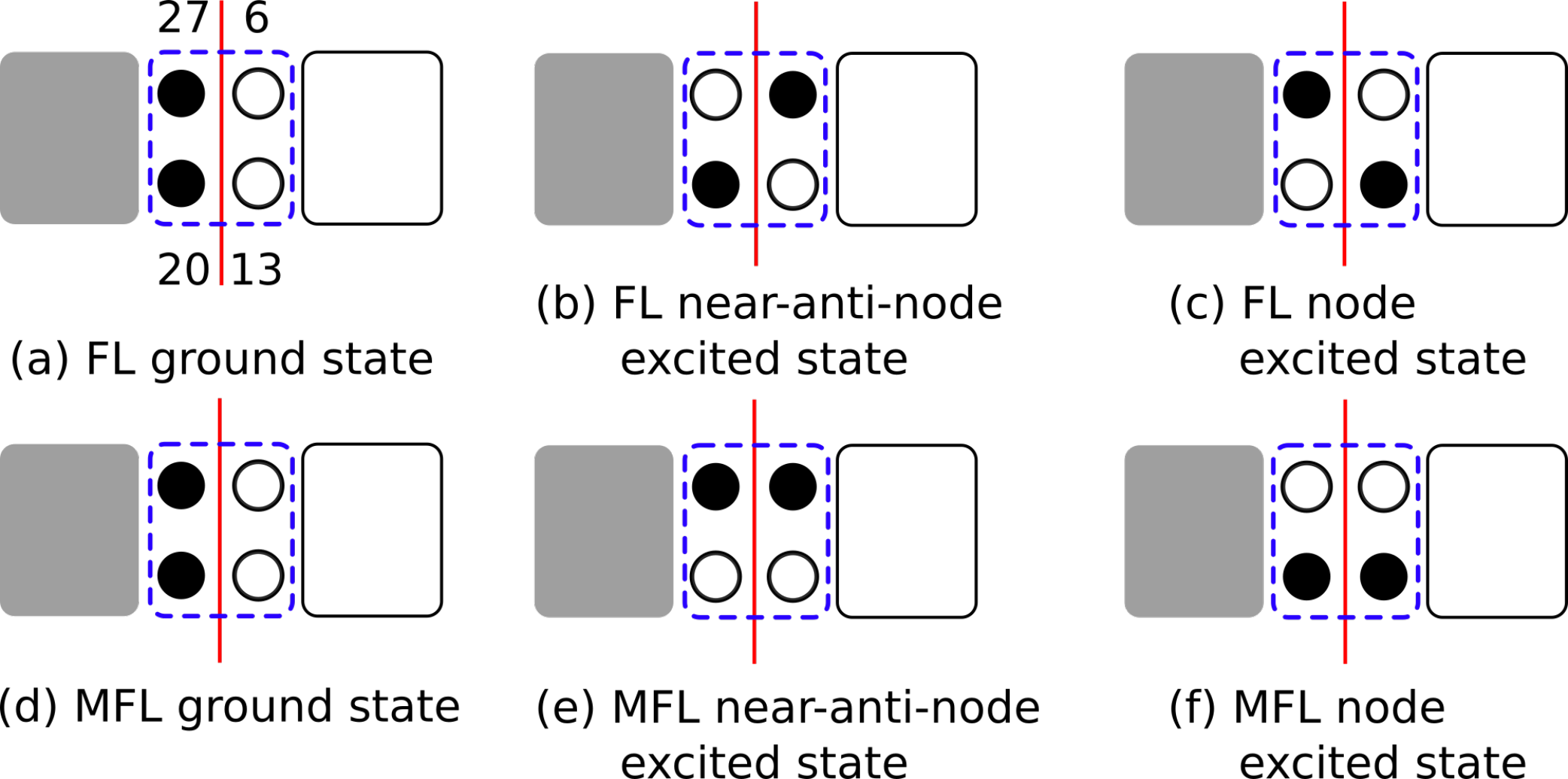}
\caption{\textbf{Construction of ground and lowest-lying excited states of FL and MFL.} The black circles represent up-pseudospins (pairs of occupied electronic Fock states), and the white circles represent down-pseudospins (pairs of unoccupied electronic Fock states). The red color vertical line in the middle represents the Fermi surface, while the grey and white boxes represent the occupied and unoccupied states below and above the Fermi surface respectively. The blue dashed box represents the window of the IR gapless quantum liquid. \textbf{(a)} FL ground state configuration: all states occupied inside the Fermi surface, and unoccupied outside. \textbf{(d)} MFL ground state configuration (identical to that of the FL). We have chosen two different excited states for both the FL and MFL phases. \textbf{(b)} FL excited state, with excitation in the near-anti-nodal direction. \textbf{(c)} FL excited state, with excitation in the nodal direction. \textbf{(e)} MFL excited state, with excitation in the near-anti-nodal direction. \textbf{(f)} MFL excited state, with excitation in the nodal direction.}
\label{fig:28_states}
\end{figure}
\pin
The IR wavefunctions for the ground state and lowest-lying excited states of the reduced system taken for the FL and MFL phases in shown in Fig.\ref{fig:28_states}. The IR fixed-point ground state involving the four Fock states ($6,13,20,27$) are identical for both the FL and MFL (see Fig.\ref{fig:28_states}(a, d)): all state below the Fermi surface are occupied, and those above it are unoccupied. This is a consequence of the fact that both these gapless liquids have been established as satisfying Luttinger's theorem~~\cite{luttinger1960ground,oshikawa2000topological,dzyaloshinskii2003some,seki2017topological, Heath2020,MukherjeeNPB1,MukherjeeTLL}. The lowest lying excited states of the FL and MFL involve configurations with holes below the Fermi surface and occupied states above it. We consider two such possible excited states in the reduced model: the near-anti-node (Fig.\ref{fig:28_states}(b) for the FL and (e) for the MFL) and nodal (Fig.\ref{fig:28_states}(c) for the FL and (f) for the MFL) excited states. 
The IR ground state of the FL is written in terms of the pseudospin degrees of freedom as $|\Psi^{FL}_{0}\rangle=|\phi^{FL}_{E}\rangle \otimes |\Psi_{IOM}\rangle_{(0)}$, where $|\phi^{FL}_{E}\rangle=|\uparrow_{20}\uparrow_{27}\rangle \otimes |\downarrow_{6} \downarrow_{13}\rangle$ represents the window of Fock states ($6,13,20,27$) proximate to the Fermi surface and the configuration of the Fock states lying outside these windows (whose occupation numbers correspond to integrals of motion (IOMs) of the URG method) is given by
\begin{eqnarray}
|\Psi_{IOM}\rangle_{(0)}= \displaystyle\prod_{i=0}^{5} |\downarrow_{i}\rangle ~ \otimes \displaystyle\prod_{i=7}^{12} |\downarrow_{i}\rangle~\otimes \displaystyle\prod_{i=14}^{19} |\uparrow_{i}\rangle \otimes \displaystyle\prod_{i=21}^{26} |\uparrow_{i}\rangle~.~~
\end{eqnarray}
\pin
The first step of the MERG then couples the four Fock states ($5,12,19,26$) with $(6,13,20,27)$ via the application of the unitary operator $\mathcal{U}_{(6)}^{\dagger}$:~$|\Psi^{FL}_{1}\rangle = \mathcal{U}_{(6)}^{\dagger}|\Psi^{FL}_{0}\rangle$~, and so onwards till we obtain a UV wavefunction that is unitarily connected to the IR state we started from.

\pin
We then employ the UV wavefunction to study the scaling of the von-Neumann entanglement entropy of a subsystem in $k$-space with its size. This is done by first selecting a block in $k$-space of size $(2\Lambda)$, chosen symmetrically about the Fermi surface and along the $\hat{s}_0$ and $\hat{s}_1$ directions. Small $k$-space blocks contain degrees of freedom corresponding to the largest real-space lengthscales 
and vice versa. By tracing over all degrees of freedom except those within the chosen block, a reduced density matrix is then obtained for the block ($\rho_{\Lambda}$). The von-Neumann entanglement entropy ($S_{EE}= -{\textrm Tr}\left[\rho_{\Lambda}\ln \rho_{\Lambda}\right ]$) is computed for the $k$-space block. Understanding the entanglement signatures in the UV wavefunctions of FL and MFL $(|\Psi_{j}\rangle,~j>0)$ is the primary goal of this work.

\begin{figure*}[!thb]
\centering
\includegraphics[width=1\textwidth]{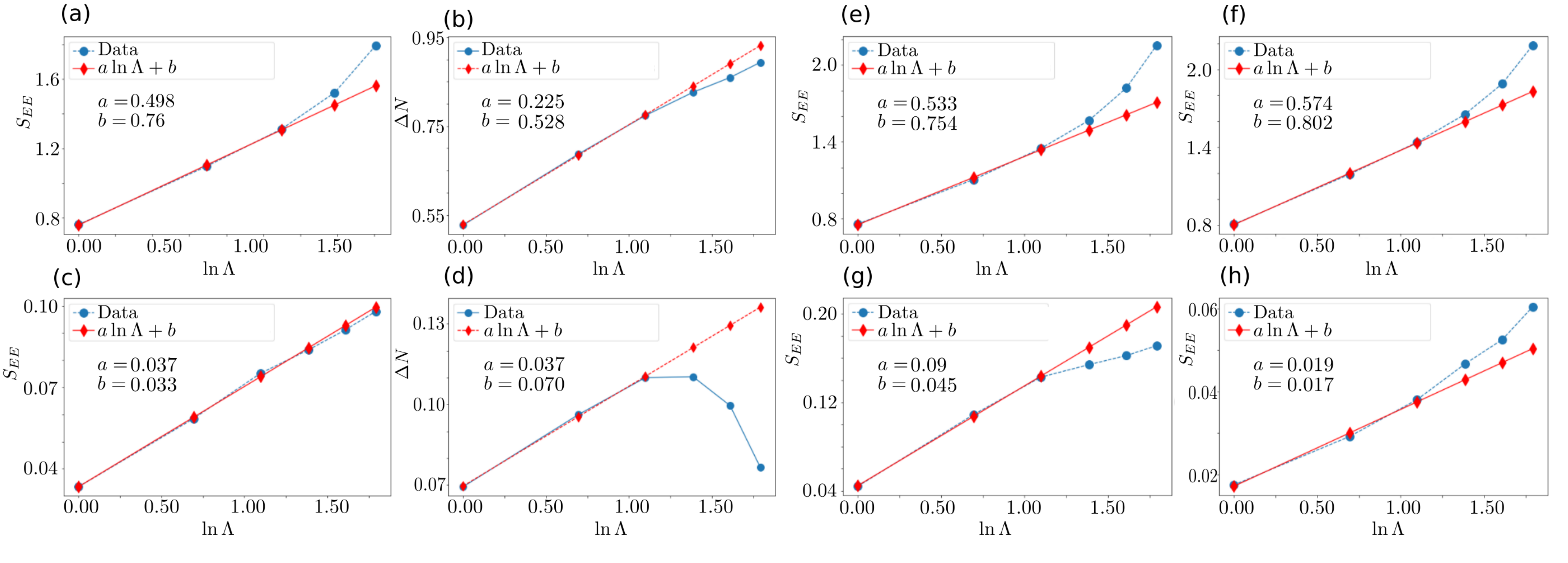}
\caption{\textbf{Scaling of entanglement entropy ($S_{EE}$) and number fluctuations ($\Delta N$) within a window in $k$-space with window size ($\Lambda$) for gapless liquids.} \textbf{(a)} $S_{EE}$ and \textbf{(b)} $\Delta N$ for FL ground state along the nodal direction $\hat{s}_{0}$. 
\textbf{(c)} $S_{EE}$ and \textbf{(d)} $\Delta N$ for MFL ground state along the nodal direction $\hat{s}_{0}$.
\textbf{(e)} and \textbf{(f)} show scaling of $S_{EE}$ for excited states of FL along nodal ($\hat{s}_0$) and antinodal ($\hat{s}_1$) directions respectively. Similarly, \textbf{(g)} and \textbf{(h)} show scaling of $S_{EE}$ for excited states of MFL along nodal ($\hat{s}_0$) and antinodal ($\hat{s}_1$) directions respectively. While $four$ MERG steps where performed for case (a), $five$ were performed in all other cases (b,e,f,c,d,g,h).
}
\label{fig:MES_NF_FL_MFL}
\end{figure*}

\pin
\textbf{\textit{Scaling of $S_{EE}$ for gapless quantum liquids}~:}~ Fig.\ref{fig:MES_NF_FL_MFL}(a) and Fig.\ref{fig:MES_NF_FL_MFL}(c) show the scaling of $S_{EE}$ $k$-space block size ($\Lambda$) along the nodal direction of the UV ground states of the FL and MFL obtained from the optimally doped and overdoped 2D Hubbard model respectively. Both gapless quantum liquids show logarithmic scaling of block entanglement in $k$-space with decrease in block size $\Lambda$, i.e., upon focusing on quantum fluctuations lying proximate to the Fermi surface:~$S_{EE}\sim \ln\Lambda$. 
The logarithmic scaling is clear for a smaller $\Lambda$ window near the Fermi surface, and deviations away from the logarithmic scaling are observed as $\Lambda$ is increased to large values. The deviations are, however, different for various types of phases. The momentum space entanglement signature suggests a corresponding logarithmic scaling in real-space as well (in large system size limit), with deviations for for small and intermediate real-space scales \cite{Ju2012,Rogerson2022}.


Further, the number fluctuations ($\Delta N=\sqrt{<N^2>-<N>^2}$) within the block $\Lambda$ also show similar logarithmic scaling for both the FL and MFL:~$\Delta N\sim\ln\Lambda$, as shown in Fig.\ref{fig:MES_NF_FL_MFL}(b) and Fig.\ref{fig:MES_NF_FL_MFL}(d) respectively. 
These results correspond to a modified area-law for these gapless liquids (upon accounting for all directions normal to the Fermi surface):~$S_{EE}^{gapless}\sim \Lambda\ln\Lambda$. Further, this confirms the existence of long-range entanglement in $k$-space encoded within the UV wavefunctions of both these gapless quantum liquids, arising from long-wavelength quantum fluctuations at low energies. 

\par
\begin{figure}
\centering
\includegraphics[width=0.49\textwidth]{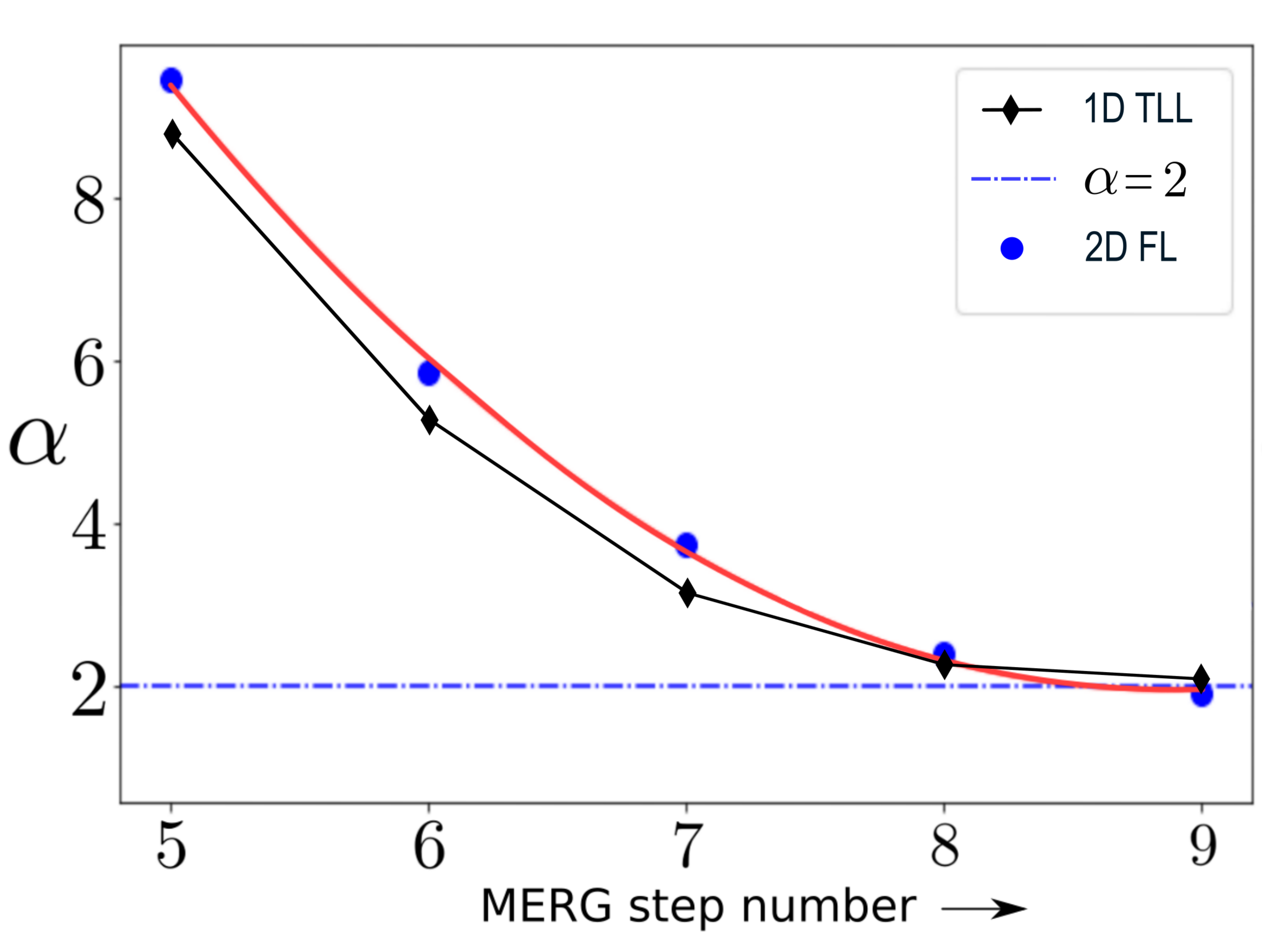}
\caption{\textbf{Variation of the inverse slope ($\alpha$) of the scaling of $S_{EE}$ with $\ln\Lambda$ with number of MERG steps.} The coefficient $\alpha$ is observed to saturate to $\alpha\sim 2$ for the Fermi liquid in two dimensions with a circular Fermi surface (blue circles), and the Tomonaga-Luttinger liquid in one dimension (black diamonds).}
\label{fig:1DFL_slope}
\end{figure}

 
The results for the scaling of $S_{EE}$ with $\Lambda$ for the two excited state configurations (i.e., along the nodal ($\hat{s}_{0}$) and near-anti-nodal ($\hat{s}_{1}$) directions normal to the Fermi surface) of the FL and MFL phases are shown in Figs.\ref{fig:MES_NF_FL_MFL}(e), \ref{fig:MES_NF_FL_MFL}(f) and Figs.\ref{fig:MES_NF_FL_MFL}(g), \ref{fig:MES_NF_FL_MFL}(h) respectively. While all results show a logarithmic scaling for $S_{EE}$ with $\Lambda$, $S_{EE}\sim (1/\alpha)\ln\Lambda$, the values of the coefficient $\alpha_{FL}$ are similar for the two excited states of the FL ($\alpha_{FL}\sim 2$). This likely arises from the presence of strong forward scattering normal to the Fermi surface, as well as tangential scattering between the two $\hat{s}$ directions. On the other hand, the values of $\alpha_{MFL}$ are quite different for the two excited states of the MFL, and $\alpha_{MFL}>>\alpha_{FL}$. This displays the strong electronic differentiation among the $\hat{s}$ directions in the MFL phase, and the fact that only strong forward scattering processes determine the entanglement of this gapless liquid. Further, we have also found that both the 2D Fermi liquid with circular Fermi surface and the 1D Tomonaga-Luttinger liquid show logarithmic scaling behaviour of $S_{EE}$ with $\Lambda$. In Fig.\ref{fig:1DFL_slope}(a) and (b), we show that the coefficient $\alpha\to 2$ for both these gapless quantum liquids upon computing the $S_{EE}$ from wavefunctions after a sufficiently large number of MERG steps. These results suggest that while the Fermi liquid and several non-Fermi liquids show the same modified area-law for the scaling of $S_{EE}$ with $\Lambda$, the value of the proportionality constant $1/\alpha$ is sensitive to the nature of the underlying Fermi surface.

\begin{figure*}[!thb]
\includegraphics[width=1\textwidth]{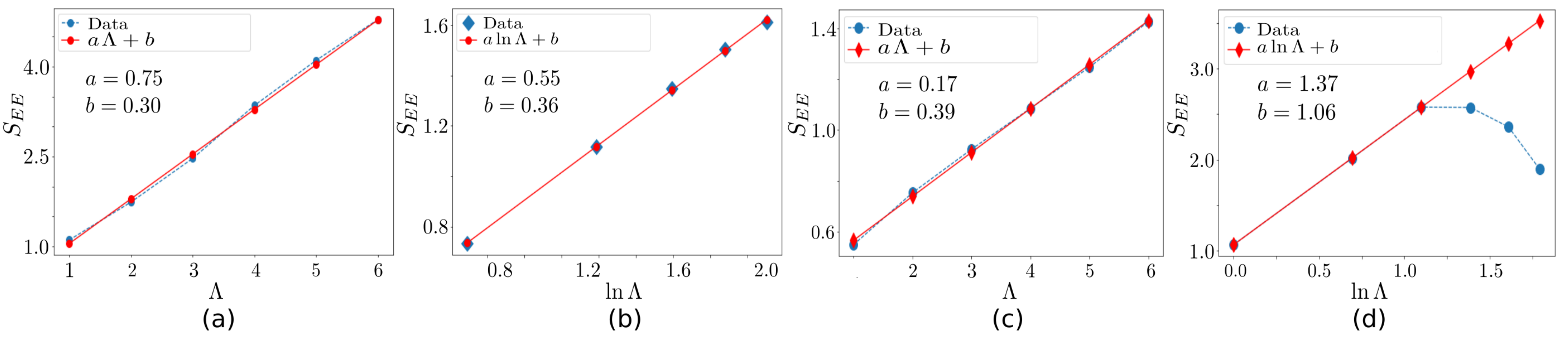}
\caption{\textbf{Scaling of entanglement entropy ($S_{EE}$) with window size ($\Lambda$) for gapped liquids.} \textbf{(a)} 2D Mott liquid along nodal ($\hat{s}_0$) direction of a 2D nested Fermi surface. \textbf{(b)} 2D Cooper pair insulator (CPI) along a radius of the circular Fermi surface. \textbf{(c)} and \textbf{(d)}: Scaling of $S_{EE}$ with $\Lambda$ for gapless MFL along nodal direction ($\hat{s}_{0}$) and 2D CPI along near-anti-nodal direction ($\hat{s}_{1}$) at the quantum critical point of the optimally hole-doped 2D Hubbard model~\cite{MukherjeeMott2} respectively. While $four$ MERG steps where performed for case (b), $five$ were performed in all other cases (a,c,d). 
}
\label{fig:MES_instabilities}
\end{figure*}

\pin
\textbf{\textit{Scaling of $S_{EE}$ for gapped quantum liquids}~:}~In order to understand whether $k$-space entanglement is a good diagnostic of fermionic criticality, we need to study RG relevant quantum fluctuations of various kinds. Having accounted above for the forward and tangential scattering related quantum fluctuations that lead to gapless quantum liquids, we now turn to the gapped quantum liquids that arise from backscattering. The ground state wavefunctions for the states studied below have been obtained from Refs.\cite{MukherjeeMott1,MukherjeeMott2,MukherjeeTLL,siddharthacpi}, and adopted to the reduced system shown in Figs.\ref{fig:cooperchannel} and \ref{fig:RGwindow} similarly to that discussed earlier for gapless liquids. 
Fig.\ref{fig:MES_instabilities}(a) shows the linear scaling of $S_{EE}$ with $\Lambda$ for the gapped 2D Mott liquid along the nodal direction $\hat{s}_{0}$. 
As we now argue, this can be understood as an area-law scaling in momentum space. The nature of the effective Hamiltonian for the 2D Mott liquid~\cite{MukherjeeMott1} and 2D Cooper pair insulator~\cite{MukherjeeMott2} obtained from instabilities of a strongly nested Fermi surface involves a strong coupling between all the electronic degrees of freedom within the gapped low-energy window lying proximate to the Fermi surface. As a result, the ground state wavefunction of these gapped quantum liquid phases cannot be decomposed into a direct product of states along each s-hat direction normal to the Fermi surface~\cite{MukherjeeMott1,ehlers2015}, leading to the area-law momentum space scaling of $S_{\textrm{EE}}\sim \Lambda$ for the 2D Mott liquid observed in Fig.\ref{fig:MES_instabilities}(a). 

On the other hand, the momentum space scaling of $S_{\textrm{EE}}\sim \ln\Lambda$ (with a proportionality coefficient $\alpha\sim 2$) for a gapped 2D Cooper pair insulator phase along a given direction normal to a circular Fermi surface observed in Fig.\ref{fig:MES_instabilities}(b) likely arises from it's non-nested nature~\cite{siddharthacpi}, while the area law scaling (similar to Fig.\ref{fig:MES_instabilities}(a)) is expected for the Cooper pair insulator arising from the strongly nested 2D Fermi surface~\cite{MukherjeeMott2}.
Further, the nature of the parent metal (from which these gapped phases are emergent) also appears to be an important factor: the MFL is the parent metal for the 2D Mott liquid~\cite{MukherjeeMott1,MukherjeeNPB2}, while the FL is the parent meal for the CPI studied in Fig.\ref{fig:MES_instabilities}(b)~\cite{siddharthacpi}. The Mott liquid and Cooper pair insulating phases are special types of gapped quantum liquids comprised of pseudo-spin degrees of freedom.

\pin In order to understand the dependence of these scaling laws of $S_{EE}$ on the nature of the underlying Fermi surface, we studied the case of the quantum critical point obtained recently in the optimally-doped 2D Hubbard model~\cite{MukherjeeMott2}: here, the four nodal directions contain gapless MFL liquids while the rest of the Fermi surface is composed of a gapped 2D Cooper pair insulator liquid.    
Thus, in Figs.\ref{fig:MES_instabilities}(c) and (d), we present the scaling of $S_{EE}$ with $\Lambda$ for the nodal MFL (along $\hat{s}_{0}$) and near-anti-nodal CPI liquid (along $\hat{s}_{1}$) respectively. While the CPI again shows the logarithmic scaling behaviour (albeit with a coefficient $\alpha<1$), the MFL shows linear scaling of $S_{EE}$ with $\Lambda$ and reflects on an enhanced long-range entanglement along the nodal directions due to an underlying quantum critical Fermi surface.

\section{Discussions}

\pin
\textbf{\textit{A modified area-law in $k$-space.}}~By studying a reduced subspace containing the relevant $k$-space scattering processes (i.e., along and between certain directions normal to the Fermi surface), our results reveal a modified area-law scaling in both the block entanglement ($S_{EE}$) and number fluctuations ($\Delta N$) for the UV ground states and lowest lying excited states unitarily connected to the Fermi liquid (FL), marginal Fermi liquid (MFL) and Tomonaga-Luttinger liquid (TLL) gapless quantum liquids (Figs.\ref{fig:MES_NF_FL_MFL} and \ref{fig:1DFL_slope})~:~$S_{EE}\propto\log (\Lambda/\Lambda_{F})$ and $\Delta N\propto\log (\Lambda/\Lambda_{F})$~(where $\Lambda_{F}$ is the Fermi momentum, and which we set to unity hereafter).
This is in agreement with findings for the real-space block entanglement entropy and number fluctuations computed from the IR state of the FL, and reveals that the forward and tangential scattering related RG relevant quantum fluctuations near the Fermi surface encode the long-range $k$-space entanglement of the modified area-law even at UV scales. This is consistent with the conjectured duality mapping between the real-space and $k$-space subsystem entanglement entropies of a system of non-interacting fermions in $d$-spatial dimensions~\cite{lee2014}. Further, it shows that the $k$-space quantum fluctuations of the UV theory are systematically converted into their real-space counterparts along the URG flow. 
\pin
\textbf{\textit{Universality of Gapless Liquids.}}~Importantly, the similarity of the results obtained for both FL, MFL and TLL offer striking evidence on the universality of 
the entanglement properties of various gapless quantum liquids that arise from systems of interacting electrons. This suggests a unification of the Fermi liquid paradigm with non-Fermi liquid metals that are qualitatively different in terms of their low-lying excitations. Interestingly, we find that the proportionality constant $\alpha$ in the relation $S_{EE}\propto (1/\alpha)\ln\Lambda$ is very different for the different metallic systems: $\alpha\sim 2$ for the FL (whether obtained from a 2D Fermi volume with a $U(1)$ symmetric circular Fermi surface (Fig.\ref{fig:1DFL_slope}(a)), or from the strongly rounded $C(4)$ symmetric Fermi surface of the FL obtained in the 2D Hubbard at large hole-doping~\cite{MukherjeeMott2} (Fig.\ref{fig:MES_NF_FL_MFL}(a), (b), (e) and (f))) 
and the TLL (with a 2 point Fermi surface, Fig.\ref{fig:1DFL_slope}(b)). On the other hand, we find $\alpha >>1$ for the MFL obtained 
in the 2D Hubbard at optimal hole-doping~\cite{MukherjeeMott2} (Fig.\ref{fig:MES_NF_FL_MFL}(c), (d), (g) and (h))). We believe that this difference arises 
from the fact that the MFL is governed by strongly RG relevant forward scattering in directions normal 
to the Fermi surface, and that MFL states on different normals show variations arising from the differentiation in the electronic dispersion everywhere along the $C(4)$ symmetric Fermi surface of the 
2D square lattice~\cite{MukherjeeMott1,MukherjeeMott2}. By contrast, the presence of subdominant RG relevant tangential scattering processes in the FL between different directions normal to the Fermi surface remove the electronic differentiation inherent in the MFL~\cite{MukherjeeMott2}.

\pin
\textbf{\textit{Classification of Gapped Quantum Liquids.}}~We also analyse the $k$-space block entanglement entropy ($S_{EE}$) of the UV states that are unitarily connected to the gapped topologically ordered quantum liquids emergent from the destabilisation of the Fermi surface of the FL and MFL gapless metals~\cite{MukherjeeMott1,MukherjeeMott2,MukherjeeNPB2,siddharthacpi}. As shown in Figs.\ref{fig:MES_instabilities}(a) and (b), we find that while the Cooper pair insulating (CPI) state for the 2D FL with circular Fermi surface~\cite{siddharthacpi} shows logarithmic scaling of $S_{EE}$ with $\Lambda$ (with the inverse coefficient $\alpha\sim 2$), the Mott liquid state in the 2D Hubbard model at $1/2$-filling~\cite{MukherjeeMott1,MukherjeeNPB2} shows an area-law scaling with $\Lambda$.
We believe that this striking difference arises from the fact that the 2D Mott liquid is emergent from the backscattering related instability of a strongly nested singular Fermi surface of the underlying 2D tight-binding model (i.e, containing van Hove singularities), while the 2D CPI state emergent from the minimal nesting of a circular Fermi surface (i.e., that of diametrically opposite Fermi points). Further, the parent metal of the former is the 2D MFL~\cite{MukherjeeMott1}, while that of the latter is the 2D FL~\cite{siddharthacpi}. Thus, we conclude that the entanglement features of a gapped quantum liquid appear to be determined by the nature of the Fermi surface and parent metal from which it emerges. Further evidence of this dependence is seen by considering the scaling behaviour of the $k$-space block entanglement entropy ($S_{EE}$) of the quantum critical state of matter found recently in the 2D Hubbard model on the square lattice at optimal doping~\cite{MukherjeeMott2}. At this novel quantum critical point (QCP), only the four nodal directions of the 2D Brillouin zone for the square lattice are found to be comprised of gapless metallic MFLs, while there exists a gapped 2D CPI quantum liquid in the antinodal regions of the rounded $C(4)$ symmetric hole-doped Fermi surface of the underlying 2D tight-binding model. We find that $S_{EE}$ scales logarithmically with $\Lambda$ for this 2D CPI state (Fig.\ref{fig:MES_instabilities}(d)), while the nodal MFLs (about four point-like singular quantum critical Fermi surfaces) show a volume-law scaling (Fig.\ref{fig:MES_instabilities}(c)). Indeed, this appears to be consistent with the findings of Ref.\cite{kaplis2017} for $S_{EE}$ of a gapless metal at a quantum critical Fermi surface. 

\pin
\textbf{\textit{Holographic evolution of entanglement.}}~Finally, the unitary renormalisation group (URG) method has also been shown to provide an explicit demonstration of the holographic principle~\cite{maldacena1999large,witten1998anti}. By this, we mean that the evolution of the Hamiltonian from UV to IR 
via the unitaries corresponds to a tensor network that admits an exact holographic mapping, i.e., it 
provides a precise relationship between the coupled degrees of freedom within one layer of the network 
with those that are emergent in the next~\cite{qi2013exact,lee2016}. Further, this generates a eigenstate 
coefficient tensor network possessing an entanglement metric~\cite{MukherjeeNPB1}. The renormalisation of 
the entanglement corresponds to the evolution of the many-particle Hilbert space geometry. By employing 
this method, we unveil several distinct holographic signatures of Fermi and non-Fermi liquids in measures of entanglement (see \ref{appendix:MBE}) and many-body correlations (see \ref{appendix:corr}). An important challenge for the future will be to characterise the entanglement properties of gapless liquids using multipartite quantum information measures, as has been achieved recently for topologically ordered gapped liquids~\cite{patra2022}. Other future studies include analysing the deviations from logarithmic scaling for any subtle signatures of universality, and investigating the microscopic origin of long-ranged entanglement in quantum critical liquid studied above as well as elsewhere~\cite{Gori2015, Ramirez2014}.
\pin 

\ack
The authors thank S. Pal, Abhirup Mukherjee, R. K. Singh, A. Dasgupta, A. Ghosh, S. Sinha and A. Taraphder for several discussions and feedback. S. P. and A. M. thanks the CSIR, Govt. of India and IISER Kolkata for funding through research fellowships. S. Lal thanks the SERB, Govt.
of India for funding through MATRICS grant MTR/2021/000141 and Core Research Grant CRG/2021/000852. We also thank two anonymous referees for their helpful comments and questions, which greatly improved our presentation.\\

\bibliography{MetalsBibliography}

\appendix

\section{Details of the URG flow equations}\label{URGdetails}
We obtain the effective Hamiltonian for the FL and MFL IR fixed point theories for the case of the half-filled square lattice by performing a URG analysis of the 2D Hubbard model~\cite{MukherjeeMott1,MukherjeeMott2}. The Hamiltonian RG flow equation is
\begin{eqnarray}
\Delta H_{(j)}=\displaystyle\sum_l \textrm{Tr}_{j,l}(c_{j,l}^{\dagger}) c_{j,l}~ G_{(j),l} ~c_{j,l}^{\dagger} \textrm{Tr}_{j,l}(H_{(j)} c_{j,l})~,
\label{eq:Delta_H}
\end{eqnarray}
where $G_{(j),l}=[\omega-\hat{n}_{j,l} \textrm{Tr}_{j,l}(H^D_{(j)}\hat{n}_{j,l}) ]^{-1}$ and $H^D_{(j)}$ is the diagonal part of the Hamiltonian at the RG step $j$. This change in the Hamiltonian is then decomposed in different scattering channels: namely, forward scattering $\Delta H^F_{(j)}$, backscattering $\Delta H^B_{(j)}$, tangential scattering $\Delta H^T_{(j)}$, and the 3-particle interaction term $\Delta H^3_{(j)}$. This leads to the following RG equations

\begin{eqnarray}
\Delta H^F_{(j)} &=& \displaystyle\sum_{k,k',l'} c_{k',l}^{\dagger} c_{k',l'}^{\dagger} \frac{4 (V(\delta)_l^{(j)})^2 \tau_{j,l} \tau_{j,l'} }{G_{j,l}^{-1}-V(\delta)_{l}^{(j)} \tau_{j,l} \tau_{j,l'}} c_{k,l'} c_{k,l}~,~~~~~~~ \textrm{where ,}~ \tau_{j,l}=(\hat{n}_{j,l}-\frac{1}{2})\\
\Delta H^B_{(j)} &=& \displaystyle\sum_{k,k',l'} c_{k',l}^{\dagger} c_{k',l'}^{\dagger} \frac{4 V(\delta)_l^{(j)} K(\delta)_l^{(j)} \tau_{j,l} \tau_{j,l'} }{G_{j,l}^{-1}-V(\delta)_{l}^{(j)} \tau_{j,l} \tau_{j,l'}} c_{k,l'} c_{k,l}~,~~~\\
\Delta H^T_{(j)} &=& \displaystyle\sum_{k,k',m,n} c_{k,m}^{\dagger} c_{k,m'}^{\dagger} \frac{ (\Gamma^{(j)})^2 (L_j^2-L_j^{z2}-L_j^z)   }{\omega - \tilde{\epsilon}^c_{j,avg} L_j^z - \Gamma^{(j)} L_j^{z2} } c_{k',n'} c_{k',n} ~,~~~\\
\Delta H^3_{(j)} &=& \displaystyle\sum_{k'',k',l',l''} c_{k',l}^{\dagger} c_{k',l'}^{\dagger} c_{j,l} \frac{V_l^{(j)}(\delta) V_l^{(j)}(\delta') \tau_{j,l}  }{\omega-\tilde{\epsilon}_{j,l}\tau_{j,l} } c_{j,l''}^{\dagger} c_{k'',l} c_{k'',l'}\nonumber\\ 
&&+   \displaystyle\sum_{k'',p',l',l''} c_{p',l}^{\dagger} c_{p',l'}^{\dagger} c_{j,l'} \frac{K_l^{(j)}(\delta) K_l^{(j)}(\delta') \tau_{j,l}  }{\omega-\tilde{\epsilon}_{j,l}\tau_{j,l} } c_{j,l''}^{\dagger} c_{k'',l} c_{k'',l'} \nonumber\\
&&+   \displaystyle\sum_{\Lambda'<\Lambda_j,p',k''} c_{p',l}^{\dagger} c_{p',l'}^{\dagger} c_{j',l'} \frac{8 R^{(j)}_{l,\delta\delta''^{(j)}}  R^{(j)}_{l,\delta'\delta''^{(j)}}  \tau_{i_1} \tau_{i_2}\tau_{i_3}}{G_{j,l,3}^{-1}- R^{(j)}_{l,\delta'',\delta''}  \tau_{i_1} \tau_{i_2} \tau_{i_3} } c_{j,l''}^{\dagger} c_{k'',l} c_{k'',l'}~.
\end{eqnarray}


In the above equations, $(j,l)$ represents the state $|\vec{k}_{\Lambda_j,\hat{s}} \sigma\rangle$, $(j,l')\equiv |\vec{k}_{-\Lambda_j+\delta, T\hat{s} \sigma}\rangle$, and $(j',l)\equiv |\vec{k}_{\Lambda_j, \hat{s} \sigma}\rangle$. Further, $V(\delta)^{(j)}_l,K(\delta)^{(j)}_l$ and $\Gamma^{(j)}_l$ are the forward scattering density-density interaction, backscattering and tangential scattering couplings respectively, while $R_{\delta\delta}^{(j)}$ is the interaction coupling for the 3-particle (i.e., 2-electron-1-hole) composites objects. $\delta$ represents the off-resonant scattering momenta. The components of the pseudospin  $\vec{L}$, $L^{+}=\sum_m c_{j,m}^{\dagger} c_{j,m'}^{\dagger}$,   $L_j^z=(1/2)\sum_m (\hat{n}_{j,m}+ \hat{n}_{j,m'}-1)$ etc. follow the standard spin-algebra. Further, the electronic Fock states are labelled as $i_1: (\vec{k}_{\Lambda_j,\hat{s},\sigma}), i_2: (\vec{k}_{-\Lambda_j+\delta^{''},T\hat{s},-\sigma}), i_3: (\vec{k}_{\Lambda'\hat{s},\sigma})$. $G_{j,l,3}^{-1}=(\epsilon_{j,l}+\epsilon_{j,l'})/2-\Delta \mu_{eff}-\omega$, where $\omega$ is the quantum fluctuation scale. From eq.\eqref{eq:Delta_H}, we extract the coupling RG equations as follows
\small{\begin{eqnarray}
\Delta V_{l,\hat{s}}^{(j)}&=&-\frac{(V_{l,\hat{s}}^{(j)})^{2}}{\frac{1}{2}E_{j,\hat{s}}-\omega+\frac{1}{4}V_{l,\hat{s}}^{(j)}}~, \\
\Delta K_{l,\hat{s}}^{(j)}&=&-\frac{(K_{l,\hat{s}}^{(j)})^{2}}{\omega - \frac{1}{2}E_{j,\hat{s}} +\frac{1}{4}K_{l,\hat{s}}^{(j)}}~, \\
\Delta \Gamma_{\hat{s}}^{(j)}&=&-\frac{N_{F}^{2}(\Gamma_{\hat{s}}^{(j)})^{2}}{ \frac{1}{N'}\sum_{s}E_{j,\hat{s}}-\omega+\frac{1}{4}\Gamma_{\hat{s}}^{(j)}}~, \\
\Delta R^{(j)}_{l,\hat{s}} &=& \frac{(V_{l,\hat{s}}^{(j)})^{2}}{\omega-\epsilon_{j,\hat{s}}} + \frac{(K_{l,\hat{s}}^{(j)})^{2}}{\omega-\epsilon_{j,\hat{s}} } + \frac{(R_{l,\hat{s}}^{(j)})^2 }{\frac{1}{2} E_{j,\hat{s}}-\omega+ \frac{1}{8} R_{l,\hat{s}}}~,
\end{eqnarray} 
}
\pin where $E_{j,\hat{s}}=(\epsilon_{\Lambda_{j,\hat{s}}}+\epsilon_{\Lambda_{j,-\hat{s}}}-2\Delta \mu_{eff})$, and $\Delta \mu_{eff}=-\Delta U/2$, 
$N_F$ is the number of electronic states on the Fermi surface, and $\epsilon_{\Lambda_{j,\hat{s}}}$ is the kinetic energy of an electron lying on the direction $\hat{s}$ normal to the Fermi surface and at a momentum-space distance $\Lambda_{j}$ from the Fermi energy. It was shown in Refs.\cite{MukherjeeMott1,MukherjeeMott2} that the gapless FL is obtained in the overdoped 2D Hubbard model with the couplings $V$ and $\Gamma$ dominant under RG, while the MFL is obtained for filling ranging from zero (i.e., a $1/2$-filled system) till optimal hole-doping at sufficiently high quantum fluctuation scale $\omega$ due to the coupling $R$ being dominant under the RG flow. Similarly, at small quantum fluctuations scales proximate to the Fermi energy, the gapped 2D Mott quantum liquid is obtained due to the backscattering coupling $K$ being dominant under RG for hole-doping ranging from zero till optimal doping. 

\section{RG evolution of various measures of Entanglement}
\label{appendix:MBE}

\subsection{Many-particle Block Entanglement Entropy (MBE)}
\label{subsec:blockentr}
\begin{figure}[!thb]
\centering
\includegraphics[scale=0.38]{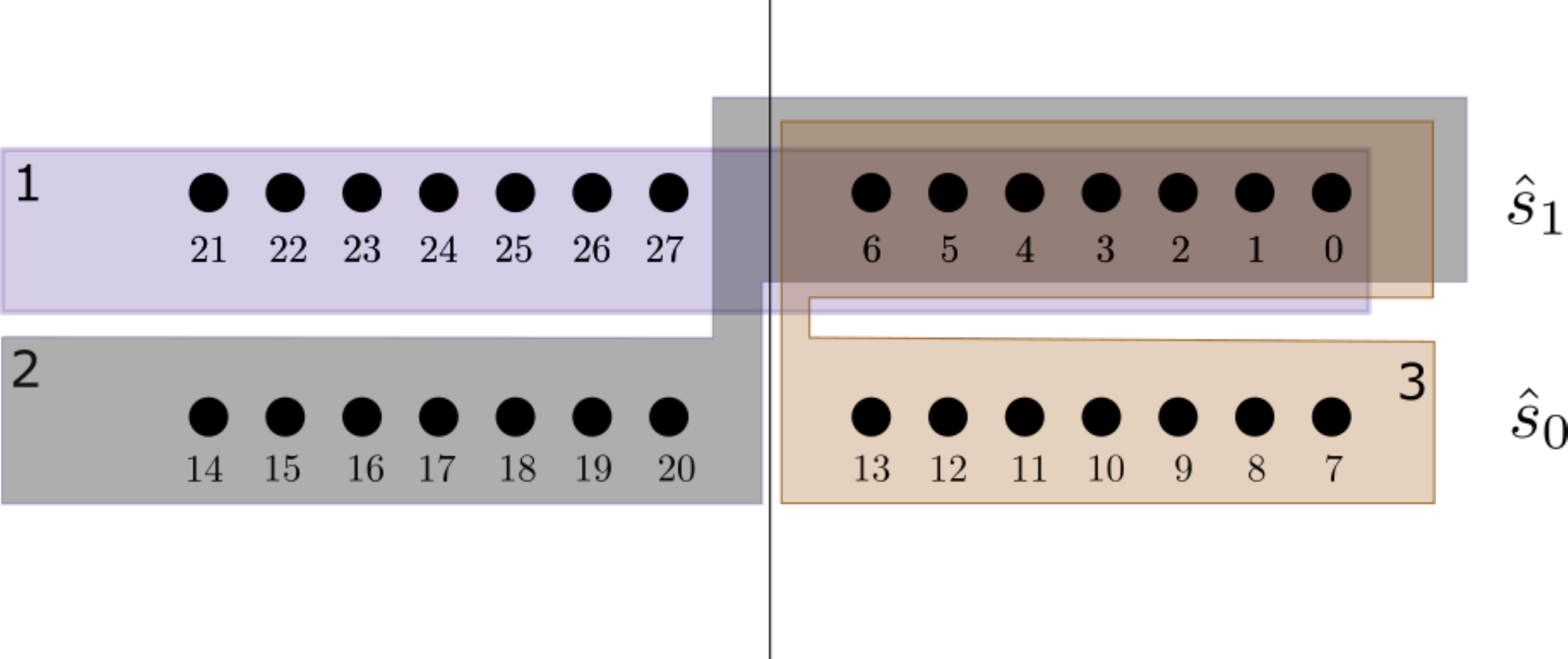}
\caption{Three choices of momentum-space blocks to compute block entanglement entropy from each member of the family of wavefunctions obtained from the MERG . \textit{Block-1} includes all the nodes on the direction $\hat{s}_1$. \textit{Block-2} includes the nodes outside the Fermi surface on $\hat{s}_1$ and inside the Fermi surface on $\hat{s}_0$. \textit{Block-3} includes all the nodes residing outside the Fermi surface.}
\label{fig:configs}
\end{figure}
\pin
As discussed in the main manuscript, we study a reduced system involving 28 pseudospin degrees of freedom formed from 56 electronic Fock states. While there is no entanglement among the low-energy  states that comprise the IR fixed point theory in momentum (or, $k$)-space, nontrivial entanglement is emergent from the MERG method. This is due to the introduction of RG-irrelevant quantum fluctuations that iteratively couple the low-energy states within the low-energy window (namely, the states $6,13,20$ and $27$) with the 24 states that lie outside it and farther away from the Fermi surface as the MERG flow proceeds towards the UV. We measure the variation of the entanglement entropy of subsystems of blocks of states in $k$-space in the many-body state with the MERG steps.
\pin
As shown in the Fig.\ref{fig:configs}, we have chosen three blocks of states in $k$-space blocks comprising a total of $14$ states. Subsystem choice $1$ consists of states along the near-anti-nodal direction $\hat{s}_{1}$ normal to the Fermi surface. Thus, the entanglement entropy measured here involves the entanglement between the states in $\hat{s}_{1}$ and those in $\hat{s}_{0})$ (i.e., the nodal direction normal to the Fermi surface), thus offering a measure of the entanglement driven by quantum fluctuations generated via tangential scatterings between the two $\hat{s}$ directions. Subsystem choice 2, on the other hand, consists of states outside the Fermi surface in $\hat{s}_{1}$ but it in $\hat{s}_{0}$; the entanglement entropy measures here the entanglement arising from quantum fluctuations involving both tangential and forward scatterings. Another subsystem choice is that shown in block 3, involving states outside the Fermi surface in both the $\hat{s}$ directions. The entanglement entropy here measures the $k$-space entanglement driven due to excitations that lead from within states the Fermi surface to those outside it, arising from quantum fluctuations involving both forward and tangential scattering events.

\begin{figure}[!thb]
\centering
\includegraphics[scale=0.28]{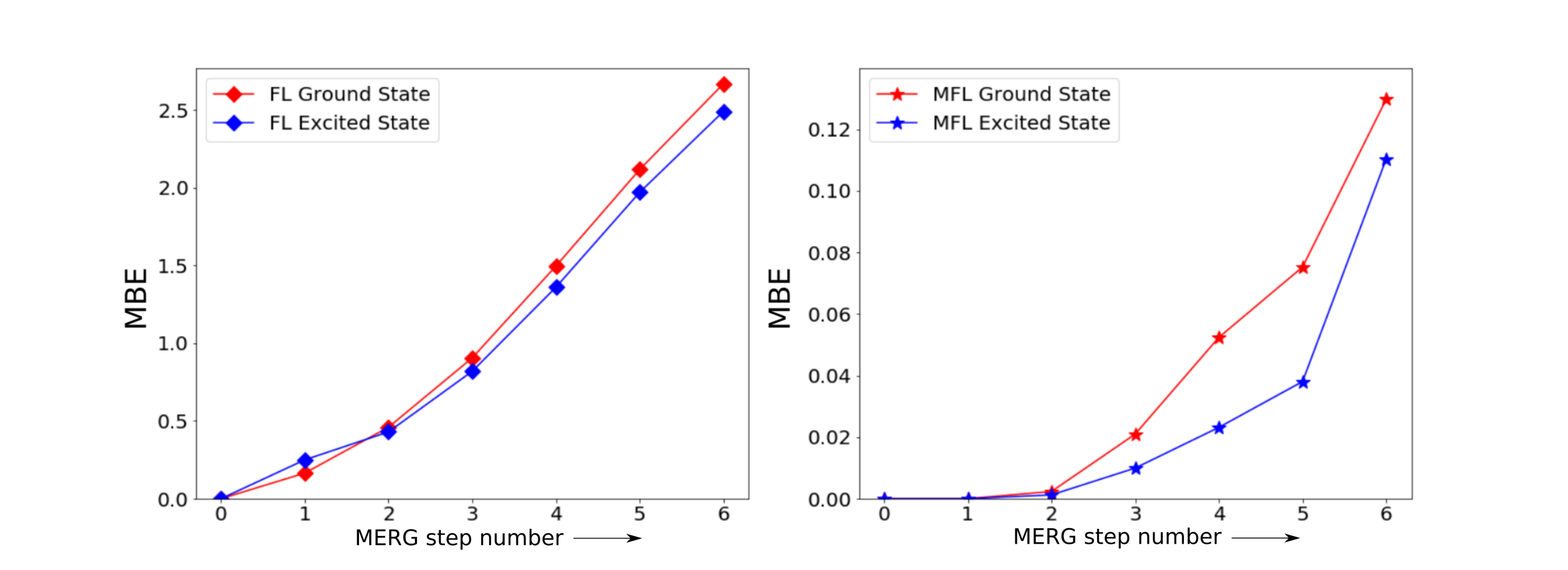}
\caption{Variation of the block entanglement entropy for the choice of Block 1 of Fig.\ref{fig:configs} with MERG step number for FL and MFL ground and nodal excited states. \textit{Left:} Red curve represents the entanglement scaling for the FL ground state, and the blue one represents the FL nodal excited state. \textit{Right:} Red curve represents the entanglement scaling for the MFL ground state, and the blue curve represents MFL nodal excited state.}
\label{fig:config1}
\end{figure}
\begin{figure}[!thb]
\centering
\includegraphics[scale=0.28]{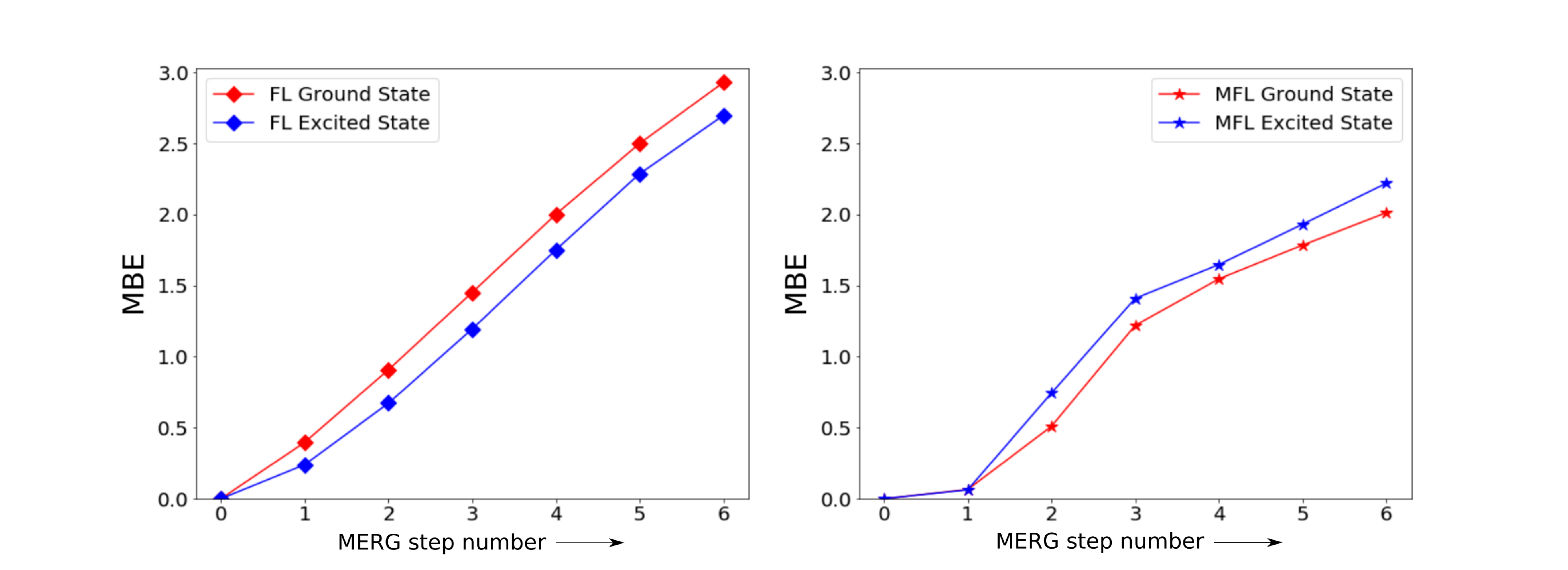}
\caption{Variation of the block entanglement entropy for the choice of Block 2 of Fig.\ref{fig:configs} with MERG step number for FL and MFL ground and nodal excited states. \textit{Left:} Red curve represents the entanglement scaling for the FL ground state, and the blue one represents the FL nodal excited state. \textit{Right:} Red curve represents the entanglement scaling for the MFL ground state, and the blue curve represents MFL nodal excited state.}
\label{fig:config2}
\end{figure}
\begin{figure}[!thb]
\centering
\includegraphics[scale=0.28]{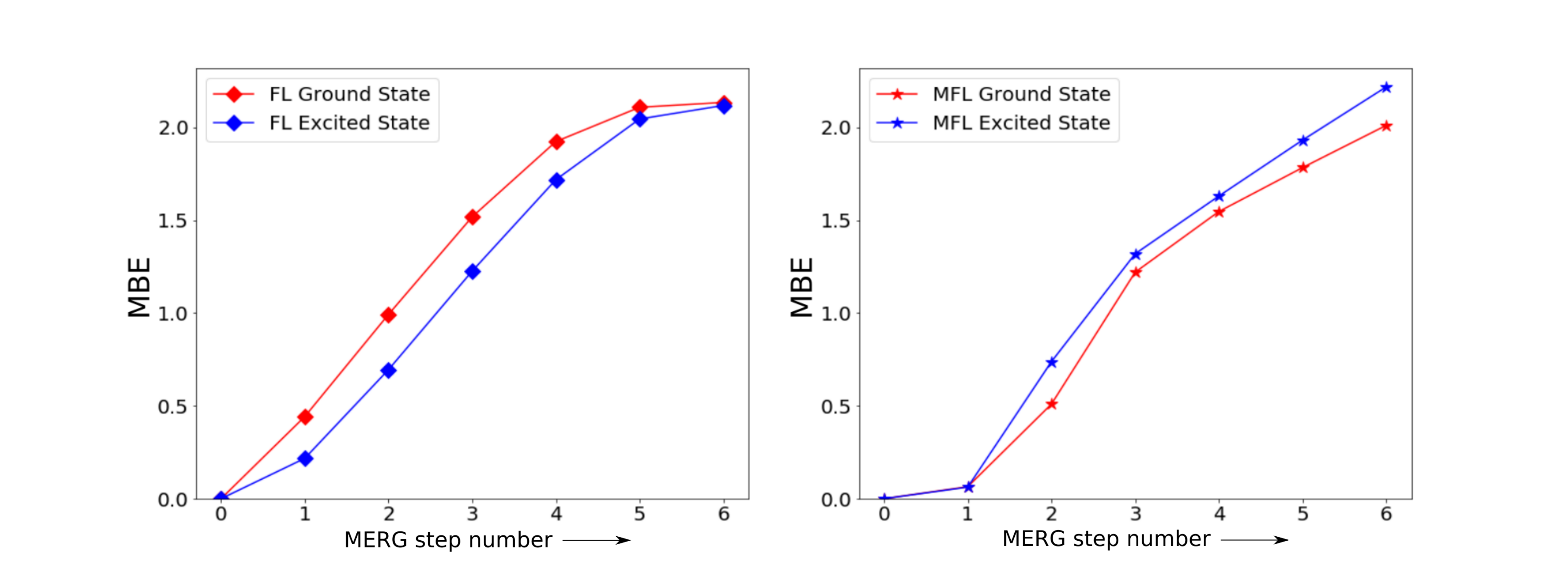}
\caption{Variation of the block entanglement entropy for the choice of Block 3 of Fig.\ref{fig:configs} with the MERG step number for FL and MFL ground and nodal excited states. \textit{Left:} Red curve represents the entanglement scaling for the FL ground state, and the blue one represents the FL nodal excited state. \textit{Right:} Red curve represents the entanglement scaling for the MFL ground state, and the blue curve represents MFL nodal excited state.}
\label{fig:config3}
\end{figure}

\pin
The results presented in Figs.\ref{fig:config1}, \ref{fig:config2} and \ref{fig:config3}) of the variation of the many-particle block entanglement (MBE) with the number of MERG steps show distinctions between the FL and MFL states. As block 1 is very sensitive to entanglement arising from quantum fluctuations due to tangential scattering events, the stark difference between the numbers achieved by the MBE in the UV for the case of block 1 in the FL and MFL reveals the dominance of tangential scattering events in the former over the latter. On the other hand, the similar values of the MBE achieved in the UV for the blocks 2 and 3 shows the importance of forward scattering events in both the FL and MFL phases. Taken together, these results show that while the IR ground state of the FL is obtained from the decoupling of quantum fluctuations that arise from both forward and tangential scattering, the MFL is obtained from the decoupling of quantum fluctuations that arise almost completely from forward scattering.

\subsection{Ryu-Takayanagi upper bound for Entanglement }
\label{subsec:ryutaka}

\begin{figure}[!h]
\centering
\includegraphics[scale=0.48]{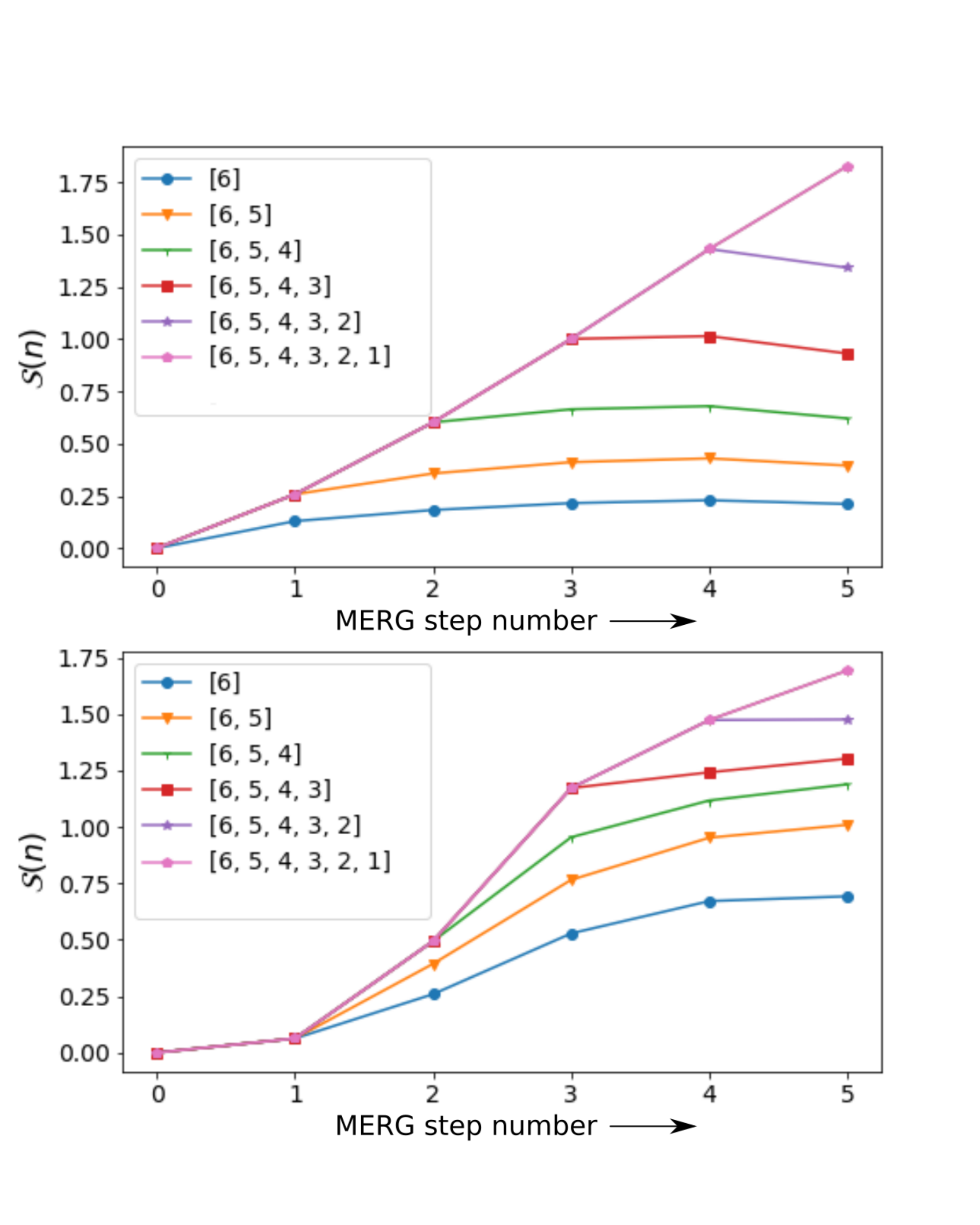}
\caption{\textit{Top:} Variation of entanglement entropy of blocks of momentum-space states of varying sizes with MERG step number for the FL ground state. Different coloured curves correspond to different block sizes (shown in inset). \textit{Bottom:} Scaling of entanglement entropy of blocks of momentum-space states of varying sizes with respect to MERG step number with MERG step number for the MFL ground state.}
\label{fig:rutak_all}				
\end{figure}
\begin{figure}[!h]
\centering
\includegraphics[scale=0.44]{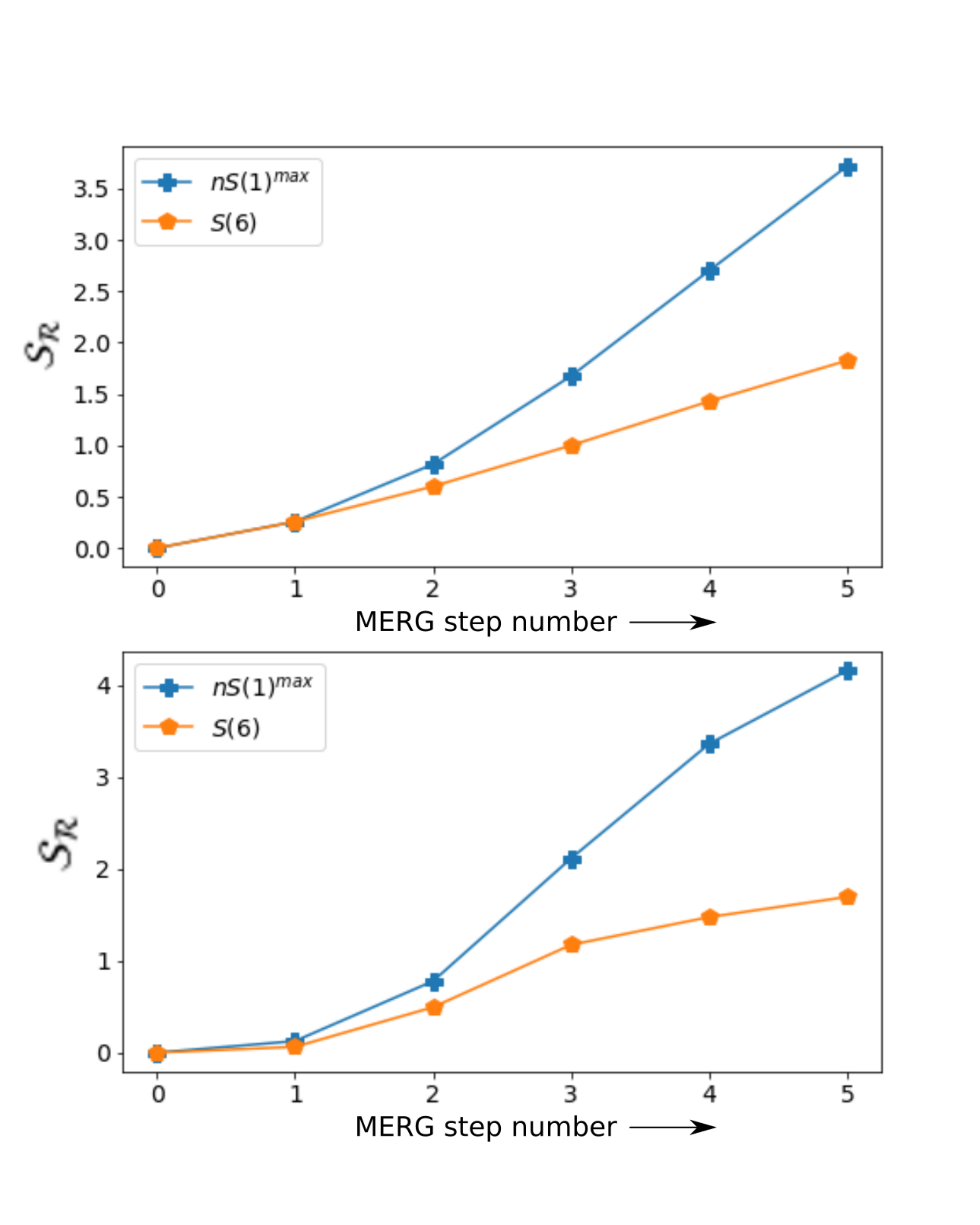}
\caption{MERG evolution of the Ryu-Takayanagi upper bound. The blue plus symbols represent the entanglement entropy of the largest block of block size six, and the red pentagons represent the upper bound by $nS(1)^{max}$, where $n=6$. The \textit{top} and \textit{bottom} figures show that the scaling of the largest block entanglement entropy is satisfied for the FL and MFL phase respectively.}
\label{fig:rutak_smax}
\end{figure}

Given that the URG method (from which the unitaries are obtained for the MERG) is known to correspond to an exact holographic mapping~\cite{qi2013exact,MukherjeeNPB1}, we will now check whether 
the entanglement entropies of various sizes of subsystems taken from the family of many-body wavefunctions generated by the MERG satisfy the Ryu-Takayanagi upper bound~\cite{RyuTaka}. This bound is defined as follows: 
\begin{eqnarray}
S^{max}_{n}\leq nS^{max}_{1}~,
\end{eqnarray}
where $S^{max}_{n}$ corresponds to the largest entanglement entropy of a subsystem containing $n$ constituents (and $n\geq 1, n\in\mathcal{Z}$).
For this, we first compute the entanglement entropy of various block sizes across the MERG energy scales ranging from the UV to the IR fixed point, as shown in the Figure.\ref{fig:rutak_all}. The blocks have been chosen to lie along the near-anti-nodal direction $\hat{s}_1$ and outside the Fermi surface. The smallest block size contains one pseudospin state ($6$), while the largest contains all $7$ states residing on the $\hat{s}_1$ direction outside the Fermi surface (see legends in inset). Fig.\ref{fig:rutak_all} shows that the block entanglement entropies increase from the IR and increase linearly in the UV for the FL (i.e., for the largest block size ($6,5,4,3,2,1,0$)), while they saturate for the MFL.  
This increase in entanglement in the FL phase is likely due to the strong growth of tangential scattering in addition to forward scattering. Fig.\ref{fig:rutak_smax} shows a comparison of $S(6)^{max}$ (associated with the largest block size $(6,5,4,3,2,1)$) for both the FL and MFL with $6\times S(1)^{max}$, and reveals that the Ryu-Takayanagi upper bound is indeed obeyed for both these gapless quantum liquids. The growth of the upper bound in the UV energy scale for the FL phase (as compared to the saturation witnessed for the MFL phase) is again likely due to the presence of strong tangential scattering processes.

\subsection{Mutual Information}
\label{subsec:mutualinformation}
\begin{figure}[!thb]
\centering
\includegraphics[scale=0.22]{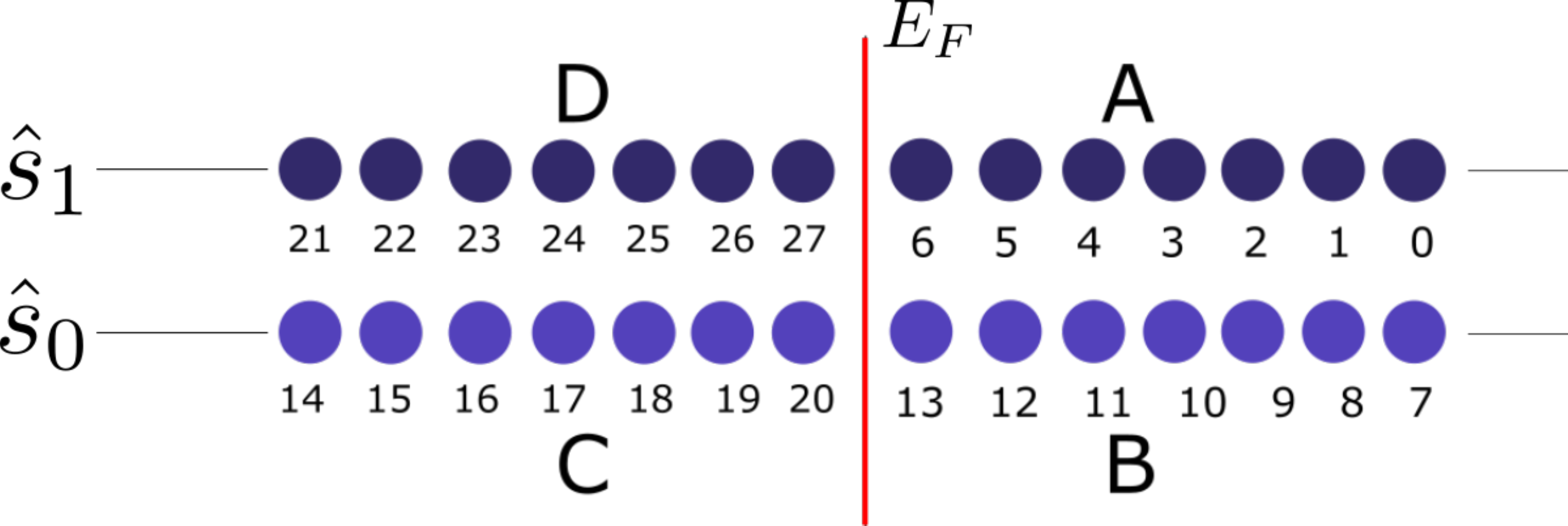}
\caption{Reduced $k$-space system of 28 pseudospins is divided into four blocks $A, B, C$ and $D$. The blocks $A$ and $B$ reside outside the Fermi surface, while the blocks $C$ and $D$ reside inside the Fermi surface. $E_F$ represents the Fermi energy. $\hat{s}_0$ and $\hat{s}_1$ represents the nodal and near-anti-nodal directions perpendicular to the Fermi surface.}
\label{fig:momentumblock}
\end{figure}
\begin{figure}[!thb]
\centering
\includegraphics[scale=0.28]{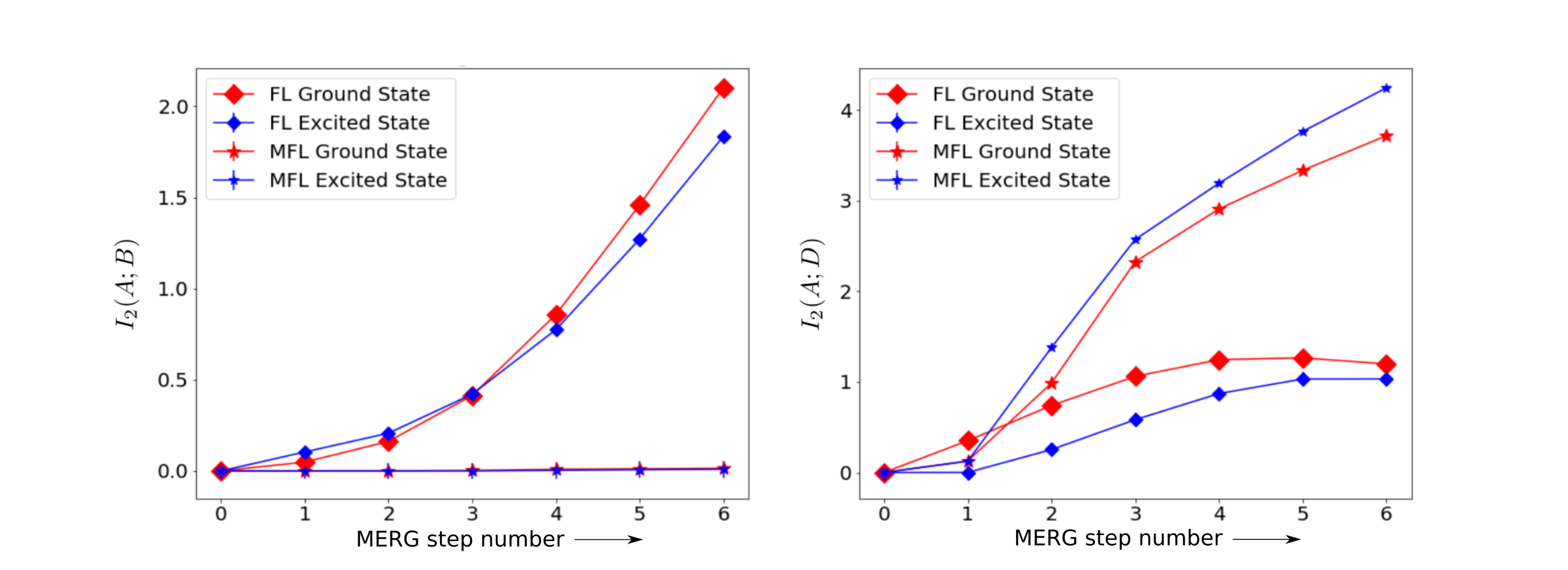}
\caption{Variation of Mutual information ($I_2(A; B)$) between blocks $A$ and $B$ (\textit{left figure}). and Mutual Information ($I_2(A; D)$) between blocks $A$ and $D$ (\textit{right figure}) with MERG step number.}
\label{fig:MI}
\end{figure}

\pin
Mutual information between two subsystems measures the total classical as well as quantum correlations between them. Definition of mutual information between two subsystems $A$ and $B$ is given as 
\begin{eqnarray}
I_2(A:B)=S(A)+S(B)-S(AB) \geq 0~,
\end{eqnarray}
where $S(A_i)$ is the von-Neumann entanglement entropy of the subset $A_i$ with the rest etc. Here, we compute the mutual information between two momentum-space blocks. For this, we divide the entire $28$ pseudospin system into four blocks ($A,B,C,D$), as shown in Fig.\ref{fig:momentumblock}, and compute the mutual information between some of them for each member of the family of wavefunctions generated by the MERG. The 
scaling of some of these mutual informations is presented in Fig.\ref{fig:MI}.
We are mainly interested in the mutual informations $I_2(A,B)$ and $I_2(A,D)$: 
which capture the correlation arising from tangential and forward scattering. Note that 
$I_{2}(A,B)$ captures the correlations between the blocks $A$ and $B$ connected via tangential and forward scattering, whereas $I_{2}(A,B)$ captures that between $A$ and $D$ connected only through forward scattering.
Fig.\ref{fig:MI}(left) shows that $I_2(A, B)$ for the ground and excited states of the FL increases with the number of MERG steps, in keeping with the steady growth in tangential scattering events as the MERG proceeds towards the UV. This is in contrast with our finding for ground and excited states of the MFL case, which remain zero throughout the MERG; this is consistent with RG irrelevant tangential scattering processes for the MFL. 
Further, the $I_{2}(A,D)$ among the blocks $A, D$ shown in Fig.\ref{fig:MI}(right) shows the importance of forward scattering is far greater in the UV for the MFL than it is for the FL.

\subsection{Tripartite Information}
\label{subsec:tripartiteInf}
\pin 
Tripartite information measures the joint entropy among three subsystems $(A, B, C)$~\cite{Kitaev_Preskill_2006,Levin_Wen_2006,patra2022}, and is defined as

\begin{eqnarray}
I_3(A,B,C)&=& S(A)+S(B)+S(C)-S(A\cup B) -S(B \cup C)-S(C \cup A) + S(A \cup B\cup C)~, \nonumber\\
&=& I_2(A,B)+S(C)-S(B \cup C)-S(C \cup A) + S(A \cup B\cup C)~. 
\label{eq:i3}
\end{eqnarray}
\begin{figure}[!thb]
\centering
\includegraphics[scale=0.24]{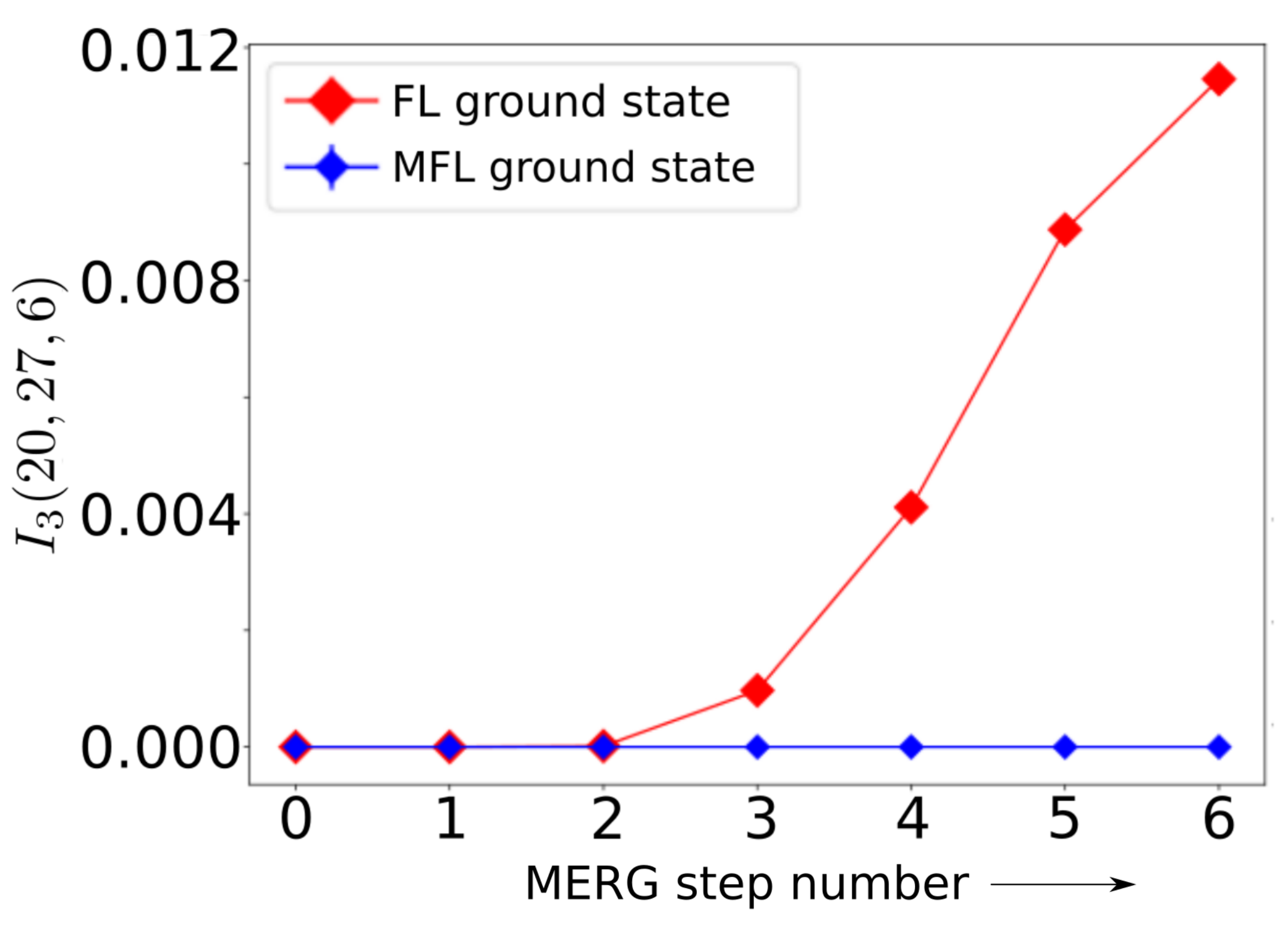}
\caption{Variation of the tripartite information $I_{3} (20,27,6)$ among the pseudospin states ($20,27,6$) with MERG step number for the FL (large red diamond) and MFL (small blue diamond) ground states.}
\label{fig:tri-mi-20276}
\end{figure} 
We are mainly interested in the tripartite information among the three pseudospin states ($20,27,6$) proximate to the Fermi surface: the states $6,27$ and $20,27$ are connected via forward and tangential scattering respectively. The MERG scaling of $I_3(6,20,27)$ for the FL and MFL ground states is shown in Figure.\ref{fig:tri-mi-20276}: while $I_3(6,20,27)$ is found to grow monotonically for the FL as the MERG proceeds to the UV, it is found to remain strictly zero throughout for the MFL. As seen above for the mutual informations, the growth of $I_3(6,20,27)$ for the FL is clearly driven by the growth in tangential scattering processes in the UV. Interestingly, the strict vanishing of $I_3(6,20,27)$ for the MFL ground state appears to show a constraint among various entanglement measures 
\begin{eqnarray}
I_3{(6,20,27)} = S(6)-S(20,6)-S(27,6)+S(20,27,6)=0~,~~~
\label{eq:mfl_constraint}
\end{eqnarray}
where we have used the fact that $I_2(20,27)=0$ in the MFL ground state. We note, however, that the individual von-Neumann entanglement entropies in eq.\eqref{eq:mfl_constraint} are nonzero.
Thus, the relations $I_2(20,27)=0$ and $I_3{(6,20,27)}=0$ appear to be invariants of the MERG flow,  capturing the decoupling between various $\hat{s}$ directions normal to the Fermi surface for the MFL.

\section{RG evolution of various many-body correlations}
\label{appendix:corr}
\subsection{Correlations generated due to pseudospin-flip scattering}
\label{subsec:spinflip}
\pin
The low-energy window is formed by the states $(6,13,2,27)$ proximate to the Fermi surface at the IR fixed point, and increases as the MERG progresses towards the UV. We define a correlation measure $C^{(j),\hat{s}_1}_{sf}$ in order to capture the pseudospin-flip correlations among the pseudospin states present in the low-energy window near the near-anti-nodal direction $\hat{s}_1$
\begin{eqnarray}
C^{(j),\hat{s}_1}_{sf}&=& \bigg( \langle \Psi^{(j)} | \displaystyle\sum^{\mathcal{U}^{\hat{s}_1}_{em}}_{i<j} (S_i^+S_j^- +  S_j^-S_j^+) | \Psi^{(j)} \rangle \bigg)/\bigg(\displaystyle\sum^{\mathcal{U}^{\hat{s}_1}_{em}}_{i<j} 1\bigg)~,~~~
\label{eq:corr_spsm}
\end{eqnarray} 
where $| \Psi^{(j)} \rangle$ is the wave function at the $j^{th}$ MERG step (i.e., $j=0$ represents the IR fixed point and increases towards the UV), $\mathcal{U}_{em}$ is the set of pseudospin states lying in the direction $\hat{s}_1$ within the low-energy window at the $j^{th}$ MERG step. (Note that $\mathcal{U}_{em}$ does not include the decoupled electronic Fock states present outside the emergent window.)
\begin{figure}[!thb]
\centering
\includegraphics[scale=0.26]{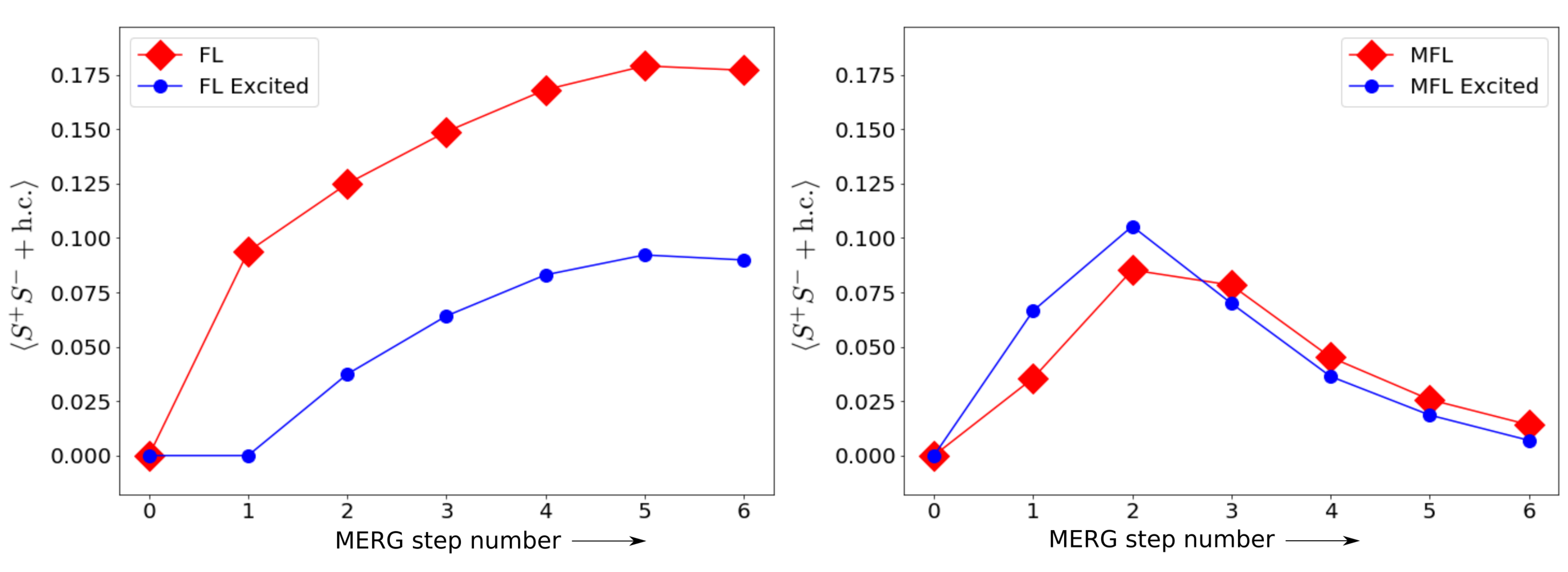}
\caption{Variation of the correlations generated by pseudospin-flip scattering processes ($\langle S^+S^- + \textrm{h.c.} \rangle$) within the low-energy window with MERG step number. \textit{Left:} Red and blue curves show variation of the correlation for the FL ground and excited state respectively. \textit{Right:} Red and blue curves show variation of the correlation scaling for the MFL ground and excited states respectively.}
\label{fig:2p_SpSm}
\end{figure}
As shown in Fig.\ref{fig:2p_SpSm}(left), the MERG scaling of the correlation $C^{(j),\hat{s}_1}_{sf}$ for the FL ground and excited states increases and finally saturates in the UV scale. This is due to the fact that the phase space for tangential scattering grows monotonically with the MERG. On the other hand, $C^{(j),\hat{s}_1}_{sf}$ correlation shows a non-monotonic behaviour for the MFL case (Fig.\ref{fig:2p_SpSm}(right)), likely arising due to an initial increase in the phase space for forward scattering at the beginning of the MERG flow and its eventual decrease as the flow reaches the UV (and highlights the absence of tangential scattering in the MFL).

\subsection{MERG scaling of the occupation of pseudospin states}
\label{subsec:numSPin}
\begin{figure}[!thb]
\centering
\includegraphics[scale=0.34]{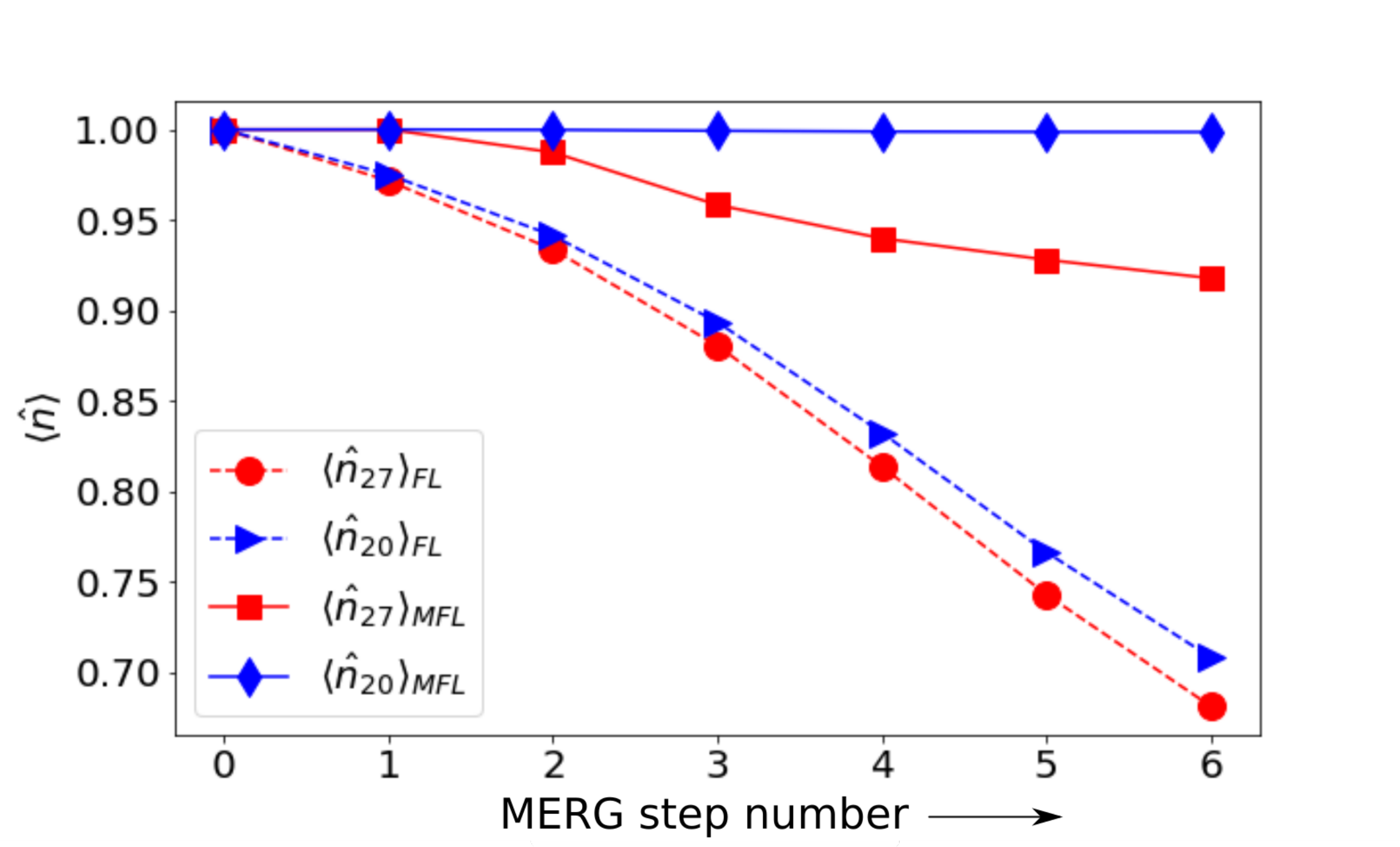}
\caption{Variation of the expectation value of the occupation of the states $20$ and $27$ with MERG step number for the FL and MFL ground states.}
\label{fig:occupation}
\end{figure}
\pin
Fig.\ref{fig:occupation} shows the MERG flow of the average number occupation of the pseudospin states $20$ and $27$ (just below the Fermi surface) for both the FL and MFL ground states shown. We see that both $\langle \hat{n}_{20}\rangle $ (blue triangle) and $\langle \hat{n}_{27}\rangle$ decay much faster for the FL (curves with blue triangles and red circles respectively) than for the MFL (curves with blue diamonds and red squares respectively). The similar nature of decays for the FL shows the weak electronic differentiation between the directions $\hat{s}_{0}$ and $\hat{s}_{1}$ due to the strong tangential scattering. On the other hand, $\langle \hat{n}_{20}\rangle$ does not decay at all under the MERG flow for the MFL ground state while $\langle \hat{n}_{20}\rangle$ decays slowly, displaying the electronic differentiation inherent in the MFL.

\subsection{Non-local strings}
\label{subsec:strings}
\begin{figure}[!thb]
\centering
\includegraphics[scale=0.50]{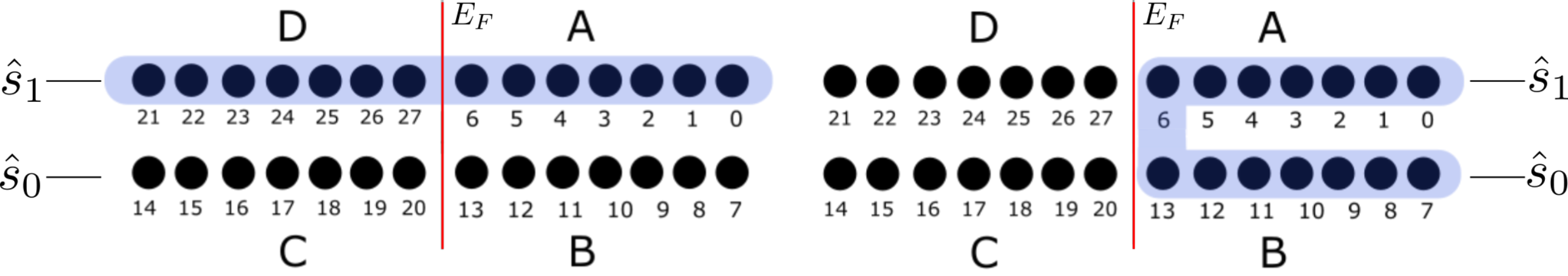}
\caption{Two choices of the non-local string parity operators. \textit{Left:} String $AD$ containing states in only one direction perpendicular to the Fermi surface ($\hat{s}_1$), and stretching across the Fermi surface. \textit{Right:} String $AB$ lying outside the Fermi surface but spread over both $\hat{s}$ directions $\hat{s}_0$ and $\hat{s}_1$.}
\label{fig:string-choice}
\end{figure}
\pin
Another way by which to probe how tangential and forward scattering shape the FL and MFL states can be learnt by computing the MERG evolution of the expectation value of the $k$-space non-local operator $Z=\prod_{k} \sigma^z_k$, $\langle Z \rangle$ for appropriately chosen strings of pseudospin states in $k$-space. As shown in Fig.\ref{fig:string-choice}, we have chosen two strings of length 14 nodes each for our analysis. 
The string $AD$ passes across the Fermi surface but along only the near-anti-nodal direction $\hat{s}_{1}$, whereas the string $AB$ lies completely outside the Fermi surface but across the two $\hat{s}$ directions. The operator $Z$ corresponds to a parity operator defined for the pseudospin states in $k$-space, i.e., sign of $\langle Z \rangle$ depends on the number of down pseudospins: negative for an odd number of down pseudospins (odd parity) and positive for an even number of down pseudospins (even parity). 
\begin{figure}[!thb]
\centering
\includegraphics[scale=0.24]{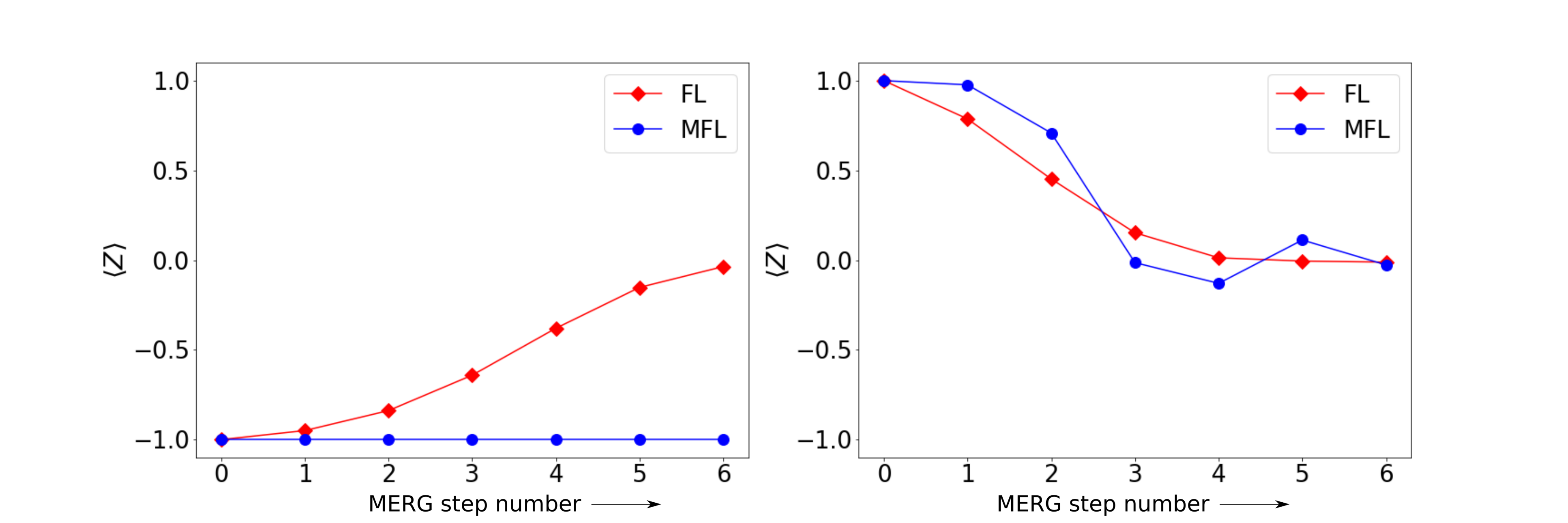}
\caption{Variation of the expectation value of the non-local string operator $\langle \hat{\mathcal{Z}} \rangle$ for two different string choices AD (left plot) and AB (right plot) (shown in Fig.\ref{fig:string-choice}) with MERG step number. The blue curve shows variation for the MFL ground state, and the red curve for the FL ground state.}
\label{fig:stringresults}
\end{figure}
First, we find from Fig.\ref{fig:stringresults}(left) that $\langle Z_{AD}\rangle$ undergoes a crossover for the FL ground state under the MERG flow from a state with a well-defined parity in the IR ($\langle Z_{AD}\rangle = -1$) to one that does not have a well-defined parity ($\langle Z_{AD}\rangle = 0$) in the UV likely due to the increased tangential scattering. On the other hand, $\langle Z_{AD}\rangle = -1$ for the MFL throughout the MERG flow, i.e., the MFL ground state maintains its parity under forward scattering. Secondly, $\langle Z_{AB}\rangle$ undergoes a crossover under the MERG flow for both the FL and the MFL from states with a well-defined parity ($\langle Z_{AB}\rangle = +1$) to one that does not have a well-defined parity ($\langle Z_{AB}\rangle = 0$). In the MFL, this arises from the absence of tangential scattering between subsystems $A$ and $B$. On the other hand, the vanishing of $\langle Z_{AB}\rangle$ for the FL likely arise from the tangential scattering events that connect states inside the Fermi surface to those outside ($D$ with $B$, and $C$ with $A$).

\subsection{Fidelity in RG scale}\label{subsec:fidelity}
\begin{figure}[!thb]
\centering
\includegraphics[scale=0.28]{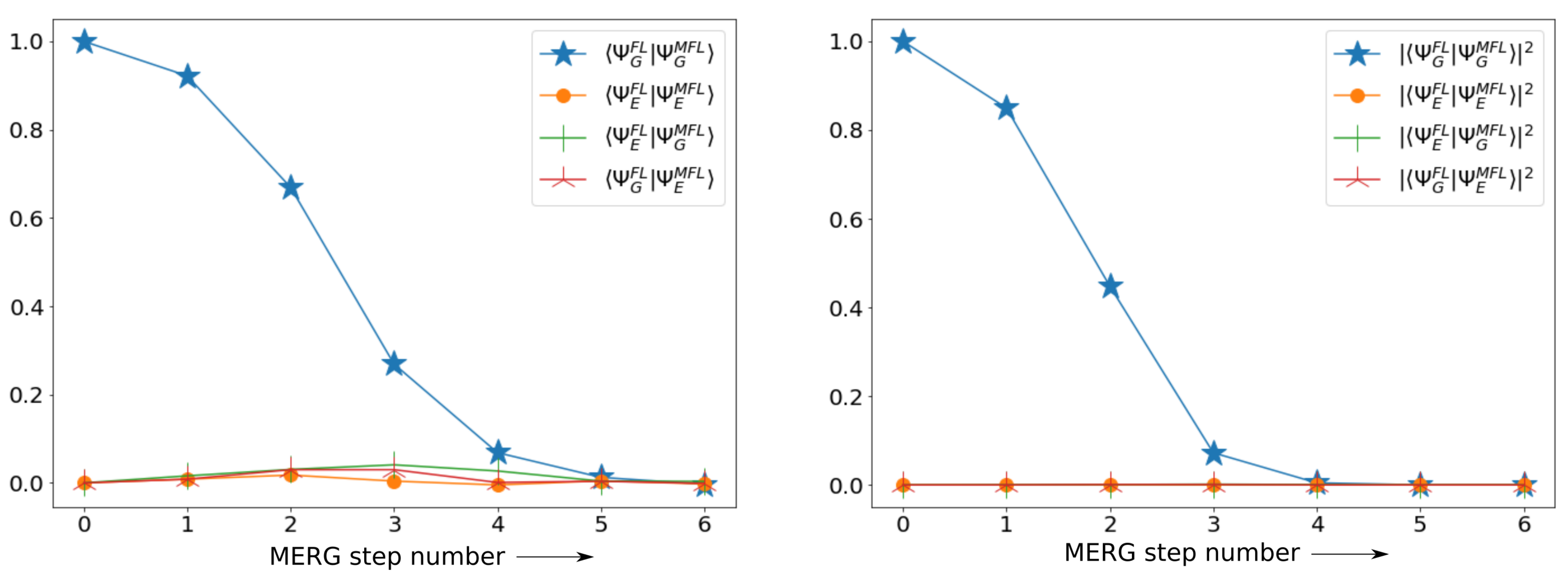}
\caption{\textit{Left:} Variation of four fidelities $F(\alpha\beta,i)=\langle \Psi^{FL}_{\alpha,0}|\Psi^{MFL}_{\beta,i}\rangle$, constructed using the overlaps of the FL and MFL ground and excited states (i.e., the indices $\alpha$ and $\beta$ can either be $G$ (ground state) or $E$ (excited state)), with MERG step number. 
\textit{Right:} This shows the square of the absolute value of the four fidelities $F(\alpha\beta,i)=\langle \Psi^{FL}_{\alpha,0}|\Psi^{MFL}_{\beta,i}\rangle$.}
\label{fig:NLL}
\end{figure}

\pin 
Fidelity is a measure of overlap between two wavefunctions defined as $F(\alpha,\beta)=|\langle \psi_{\alpha} |\psi_{\beta} \rangle|^2$. We now study the fidelity between the wavefunctions of the FL and MFL ground and excited states as the MERG proceeds from IR to UV. We define $|\Psi^{FL}_{G/E,0} \rangle$ and $|\Psi^{MFL}_{G/E,0} \rangle$ as the IR fixed-point ground ($G$) and excited ($E$) state wavefunctions for the FL and MFL phases respectively. Similarly, $|\Psi^{FL}_{G/E,i}\rangle,|\Psi^{MFL}_{G/E,i}\rangle$ are the ground ($G$) and excited ($E$) state wavefunctions for the FL and MFL at the $i^{th}$ MERG step.
Then, we define the four possible fidelities between the ground and excited state wavefunctions of the FL and MFL phases at the $0^{th}$ and $i^{th}$ MERG steps as given by $F(\alpha\beta,i)=\langle \Psi^{FL}_{\alpha,0}|\Psi^{MFL}_{\beta,i}\rangle$, where the indices $\alpha$ and $\beta$ can either be $G$ (ground state) or $E$ (excited state). 
The MERG flow of the four different fidelities $F(\alpha\beta,i)=\langle \Psi^{FL}_{\alpha,0}|\Psi^{MFL}_{\beta,i}\rangle$ is shown in Fig.\ref{fig:NLL}.
\pin
We can see from Fig.\ref{fig:NLL}(left) that the fidelity between the FL and MFL ground states is perfect at the IR fixed point, $F_{GG,i=0} = 1$. This is simply because both these gapless quantum liquids satisfy Luttinger's theorem, and have the same ground state. However, $F_{GG}$ falls monotonically to zero. On the other hand, the IR excited states of the FL and MFL are orthogonal to the common IR ground state and to one another, as indicated by the near vanishing of the other three fidelities at the zeroth MERG step. Further, this orthogonality is maintained under the MERG flow to the UV, as the unitary operations of the MERG maintain the orthogonality structure of these states. Fig.\ref{fig:NLL}(right) shows that the MERG evolution of the square of the real part of the fidelities $F(\alpha\beta,i)$ follows the same trend as $F(\alpha\beta,i)$ itself, reinforcing the fact that the phases of the four fidelities do not play an important role.

\end{document}